%% file: revised_version.tex
\documentclass[usenames,dvipsnames]{aa}
\usepackage{graphicx}
\graphicspath{{./figures/}}
\usepackage[varg]{txfonts}
\usepackage{natbib}
\bibpunct{(}{)}{;}{a}{}{,} 
\usepackage{xcolor}
\usepackage{siunitx}
\usepackage[export]{adjustbox}
\usepackage{subcaption}
\usepackage{pdflscape}
\usepackage{ulem}

\usepackage{hyperref}
\hypersetup{colorlinks=true, linkcolor=Maroon,
    citecolor=PineGreen,
    bookmarksnumbered=true, pdfborder={0 0 0},unicode,breaklinks}

\begin{document}

\title{Double detonations of sub-M$_{\text{Ch}}$ CO white dwarfs: Can different core and He shell masses explain variations of Type Ia supernovae?}

\author{Sabrina~Gronow\inst{1,2}\fnmsep\thanks{\email{sabrinagronow2@gmail.com},
Fellow of the International Max Planck Research School for Astronomy and Cosmic
Physics at Heidelberg University (IMPRS-HD)} \and
Christine~E.~Collins\inst{3,4} \and
Stuart~A.~Sim\inst{3} \and
Friedrich~K.~R\"{o}pke\inst{2,5} }

\titlerunning{Influence of different core and He shell masses}
\authorrunning{Gronow et al.}

\institute{
Zentrum f\"ur Astronomie der Universit\"at Heidelberg,
Astronomisches Rechen-Institut, M\"{o}nchhofstr. 12-14, 69120 Heidelberg, Germany
\and
Heidelberger Institut f\"{u}r Theoretische Studien,
Schloss-Wolfsbrunnenweg 35, 69118 Heidelberg, Germany
\and
Astrophysics Research Center, School of Mathematics and Physics, Queen's
University Belfast, Belfast BT7 1NN, Northern Ireland, UK
\and
GSI Helmholtzzentrum f\"{u}r Schwerionenforschung, Planckstraße 1, 64291 Darmstadt, Germany
\and
Zentrum f\"ur Astronomie der Universit\"at Heidelberg, Institut f\"ur
theoretische Astrophysik, Philosophenweg 12, 69120 Heidelberg, Germany}

\date{Received November 20, 2020} 

\abstract
{Sub-Chandrasekhar mass carbon-oxygen white dwarfs with a surface helium shell
    have been proposed as progenitors of Type Ia supernovae (SNe~Ia). If true,
    the resulting thermonuclear explosions should be able to account for at
    least some of the range of SNe~Ia observables. To study this, we conduct a
    parameter study based on three-dimensional simulations of double
    detonations in carbon-oxygen white dwarfs with a helium shell, assuming
    different core and shell masses. An admixture of carbon to the shell and
    solar metallicity are included in the models. The hydrodynamic simulations
    are carried out using the \textsc{Arepo} code. This allows us to follow the
    helium shell detonation with high numerical resolution, and improves the
    reliability of predicted nucleosynthetic shell detonation yields. The
    addition of carbon to the shell leads to a lower production of $^{56}$Ni
    while including solar metallicity increases the production of IMEs. The
    production of higher mass elements is further shifted to stable isotopes at
    solar metallicity. Moreover, we find different core detonation ignition
    mechanisms depending on the core and shell mass configuration. This has an
    influence on the ejecta structure.  We present the bolometric light curves
    predicted from our explosion simulations using the Monte Carlo radiative
    transfer code \textsc{artis}, and make comparisons with bolometric SNe Ia
    data.  The bolometric light curves of our models show a range of
    brightnesses, able to account for sub-luminous to normal brightness SNe Ia.
    We show the model bolometric width-luminosity relation compared to data for
    a range of model viewing angles. We find that, on average, our brighter
    models lie within the observed data.  The ejecta asymmetries produce a wide
    distribution of observables, which might account for outliers in the data.
    However, the models overestimate the extent of this compared to data. We
    also find the bolometric decline rate over 40 days,
    $\Delta\rm{m}_{40}$(bol), appears systematically faster than data.
}

\keywords{Hydrodynamics -- Methods: numerical -- Nucleosynthesis, abundances -- Radiative transfer -- Supernovae: general -- White dwarfs}

\maketitle

\section{Introduction}
\label{sec:introduction}
A widely discussed progenitor of a Type Ia supernova (SN Ia) is a
sub-Chandrasekhar mass (sub-M$_{\text{Ch}}$) white dwarf (WD)
\citep[e.g.][]{shigeyama1992a, nugent1997a, hoeflich1998b, garcia1999a,
fink2007a, fink2010a, sim2010a, sim2013b, blondin2017b, shen2018a, liu2018a,
polin2019a, leung2020a} in a binary system. Its detonation can reproduce
several observed features of a SN Ia \citep{sim2010a}.  Compared to the
alternative model of a WD exploding when it approaches the Chandrasekhar mass,
a strength of this scenario is that it directly relates the amount of produced
$^{56}$Ni and thus the brightness of the event to a fundamental parameter of
the progenitor -- its mass. Such explosions roughly follow the trend of the
observed ``width-luminosity'' relation between the peak brightness and decay
after maximum light in the $B$-band \citep{phillips1993a, phillips1999a}; again
with the progenitor's mass as the ordering parameter (\citealp{sim2010a},
predicted by \citealp{pinto2000a}). In contrast, models fixing the progenitor
mass to the Chandrasekhar limit struggle with reproducing this important
relation (\citealp{sim2013a}, but see \citealp{kasen2009a}, and
\citealp{blondin2017a, shen2018b} for sub--M$_{\text{Ch}}$ models).
\citet{kushnir2020b} and \citet{sharon2020a}, however, argue that the
$t_0$-$M_{^{56}\mathrm{Ni}}$ relation cannot be reproduced with
sub-M$_{\text{Ch}}$ WD models to date, with $t_0$ being the $\gamma$-ray escape
time from the ejecta which can be determined from bolometric light curves
\citep{wygoda2019a}. Therefore it is of interest to investigate the progenitor
models for normal SNe~Ia. We do this here as validation of the double
detonation scenario.

The drawback of the sub-M$_{\text{Ch}}$ scenario is that the initiation of the
thermonuclear explosion is not as straightforward as in the Chandrasekhar-mass
model \citep[see][for reviews on different progenitor systems and explosion
scenarios]{hillebrandt2013a, maoz2014a, livio2018a, roepke2018a, soker2019a}.
Several mechanisms have been proposed for initiating the detonation of a
sub-M$_{\text{Ch}}$ WD. WDs can interact with another WD (double degenerate
scenario) \citep[e.g.][]{whelan1973a, webbink1984a, kashyap2015a,
tanikawa2018a, rebassa2019a} or with another star, such as a He star, (single
degenerate scenario) in a binary system \citep[e.g.][]{whelan1973a,dave2017a}.
A violent merger of two WDs \citep{guillochon2010a, pakmor2010a, pakmor2011b,
pakmor2013a} could ignite an explosion as well. The most violent mechanism is
presented in a collision model \citep[e.g.][]{piro2014b, wygoda2019a}.

Population synthesis calculations by \citet{belczynski2005a} and
\citet{ruiter2009a} indicate that the double degenerate scenario occurs often
enough to explain a significant part of the SN Ia rate. However, studies by
other groups, such as \citet{toonen2012a} and \citet{liu2018a}, indicate that
the contribution is lower.

A much discussed explosion mechanism is the double detonation scenario
\citep[e.g.][]{woosley1994b,fink2007a, fink2010a, moll2013a, shen2018a,
townsley2019a, leung2020a, gronow2020a}. In this case, a carbon-oxygen (CO) WD
accretes helium (He) from a companion, e.g.\ a He star \citep{iben1987a} or He
WD \citep{tutukov1996a}. A He detonation can be ignited through compressional
heating of the accreted material. This detonation propagates through the He
shell and sends a shock wave into the CO core. A second detonation in the
carbon-oxygen material is ignited following its convergence.

We carry out a parameter study to investigate the effects of different core and
He shell masses of exploding WDs. \citet{neunteufel2016a} model short period
binary systems of a CO WD with a He star. They consider different WD core
masses and follow the accretion process. The mass ranges they find for the
accreted He shell (see their Figure~6) are partially covered in our parameter
study and depend on the donor mass as well as the orbital period. Similar
parameter studies to ours have been carried out by \citet{fink2007a,fink2010a},
\citet{polin2019a} and \citet{leung2020a}. \citet{fink2007a} look into the
effect of different ignition configurations in the He shell while
\citet{fink2010a} only consider certain core and shell mass combinations.
\citet{polin2019a} study a much wider parameter space which partially overlaps
with that in our work. They, however, consider zero metallicity of the zero-age
main sequence progenitor and follow the evolution with one-dimensional (1D)
explosion simulations. The models by \citet{fink2007a, fink2010a} and
\citet{leung2020a} are two-dimensional (2D). Three dimensional (3D) simulations
are carried out by \citet{moll2013a}, though they only compute one quarter of
the WD.

The parameter study presented here is based on full 3D simulations using the
moving mesh code \textsc{Arepo} \citep{springel2010a}. This approach allows for
a more accurate treatment of the detonation dynamics than the level-set method
used by \citet{fink2010a}. The mass of the CO WD cores is set to be between
$0.8$\,$M_\odot$ and $1.1$\,$M_\odot$. Since the He shell masses depend on the
accretion rate, our models consider a range of $0.02$\,$M_\odot$ to
$0.1$\,$M_\odot$. The shell and core mass combinations are chosen to match
models in previous work \citep[e.g.][]{woosley2011a,polin2019a,townsley2019a}.
Of particular interest are models with low-mass He shells \citep{fink2010a,
townsley2019a} because the imprints of massive He shell detonations are found
to be inconsistent with observations of normal SNe~Ia
\citep{hoeflich1996b,kromer2010a}. The admixture of carbon from the WD core
into the shell considered in our models further decreases the amount of free
$\alpha$-particles and less heavy elements are produced
\citep[see][]{yoon2004b,fink2010a,gronow2020a}. All our models are calculated
assuming solar metallicity of the zero-age main sequence progenitors.

The \textsc{Arepo} code enables us to increase the resolution in selected
regions. Using its adaptive mesh refinement, we reach a higher resolution in
the He shell compared to previous work \citep[e.g.][]{fink2007a,moll2013a}.
Consequently, the propagation of the He detonation wave and its critical
nucleosynthesis yields can be modeled more accurately than in previous studies.

The methods used in this parameter study are described in Section
\ref{sec:methods}. In Section \ref{sec:models} we present the model setup
succeeded by a discussion of the results from the explosion simulations in
Section \ref{sec:results}. A discussion of our models in the context of
previous simulations follows in Section \ref{sec:discussion}. Synthetic
observables are analyzed in Section \ref{sec:observables}. The conclusions are
presented in Section \ref{sec:summary}. The 1D structure and nucleosynthesis
yields will be made available on the supernova archive HESMA
\citep{kromer2017a}.

\section{Methods}
\label{sec:methods}

\subsection{Hydrodynamics}
\label{sec:hydro}
We carry out 3D simulations using the \textsc{Arepo} code
\citep{springel2010a}. Extensions, such as the Helmholtz equation of state
\citep{timmes2000a}, were implemented by
\citet{pakmor2012b,pakmor2013a,pakmor2016a}. Other extensions allow us to
couple the hydrodynamic solver of the moving mesh code to a nuclear network
solver \citep{pakmor2012b}. The energy equation and balance equations for
nuclear isotopes are extended by a source term to model reactive flows. We use
the same methods as in \citet{gronow2020a}: the \textsc{Arepo} code
\citep{springel2010a, pakmor2016a} employs a second-order finite-volume method
in combination with a tree-based gravity solver to integrate the Euler-Poisson
equations forward in time. Instead of the 33 isotope network of
\citet{gronow2020a}, we include 35 species in our nuclear network, now also
accounting for $^{14}$N and $^{22}$Ne which represent the metallicity of the WD in
the hydrodynamics simulations. It is comprised of n, p, $^4$He, $^{12}$C,
$^{13}$N, $^{14}$N, $^{16}$O, $^{20}$Ne, $^{22}$Ne, $^{22}$Na, $^{23}$Na,
$^{24}$Mg, $^{25}$Mg, $^{26}$Mg, $^{27}$Al, $^{28}$Si, $^{29}$Si, $^{30}$Si,
$^{31}$P, $^{32}$S, $^{36}$Ar, $^{40}$Ca, $^{44}$Ti, $^{45}$Ti, $^{46}$Ti,
$^{47}$V, $^{48}$Cr, $^{49}$Cr, $^{50}$Cr, $^{51}$Mn, $^{52}$Fe, $^{53}$Fe,
$^{54}$Fe, $^{55}$Co, and $^{56}$Ni. One of our models with a particularly low
He shell mass is calculated with a 55 isotope nuclear network, because in this
case an extended network is required to capture the energy release more
accurately \citep[see][for a detailed explanation]{shen2018b,townsley2019a}.
The details of the nuclear network are given in Section 4.7 of
\citet{gronow2020a}.

In our simplified treatment in the hydrodynamical explosion simulations, the
metallicity is set by adding an appropriate amount of $^{22}$Ne in the core and
$^{14}$N in the shell. The values are calculated based on solar metallicity
\citep{asplund2009a}: All initial carbon and oxygen accumulates in $^{14}$N
during CNO cycle hydrogen burning in the shell. In the core material, it is
converted to $^{22}$Ne. The composition of the core is set to the mass
fractions $X(^{12}\mathrm{C}) = 0.5$, $X(\mathrm{^{16}O})=0.49$ and
$X(^{22}\mathrm{Ne})=0.01$; for the shell we use $X(^4\mathrm{He}) = 0.997$ and
$X(^{14}\mathrm{N})=0.003$. A larger nuclear network in the hydrodynamic
simulations would significantly increase the computational costs but not affect
the dynamics of the explosions themselves. Detailed nucleosynthesis yields are
instead determined in a postprocessing step based on tracer particles
\citep{travaglio2004a} with a much larger network.

For modeling detonations on the \textsc{Arepo} grid, burning is disabled inside
the shock. A detailed description of the implementation can be found in
\citet{gronow2020a} and follows \citet{fryxell1989a} and Appendix A in
\citet{townsley2016a}. It differs from the scheme of \citet{kushnir2020a} who
introduce a burning limiter for the modeling of thermonuclear detonation waves.
The scaling factor employed in the limiting procedure is sensitive to the setup,
and a detailed calibration is necessary which goes beyond our current work.
Pakmor et al. (subm.), however, investigate the effect of the burning limiter
by \citet{kushnir2020a} in \textsc{Arepo} simulations of mergers involving
hybrid HeCO WDs. They find that their main simulation results do not dependent
on the use of the burning limiter, and therefore our simulations were carried
out with the implementation following \citet{fryxell1989a} and
\citet{townsley2016a}.

\textsc{Arepo} allows for adaptive mesh refinement. This is exploited here in
all models in the same way as in \citet{gronow2020a}: The He shell and the
region where the He detonation wave converges opposite to its ignition spot
have higher levels of refinement. The region at the antipode of the He ignition
spot was found to be critical for the detection of the carbon detonation
ignition mechanism by \citet{gronow2020a}. Since the convergence region is
located in the shell, this region has the highest level of refinement, followed
by the He shell and the remaining WD at base resolution. A cell is refined
based on where it is located. It is split when its mass exceeds a target mass
by a prescribed factor as in \citet{pakmor2013a} and \citet{gronow2020a}. A
passive scalar is placed into the He shell to follow its location. We use the
same reference mass of $2\times10^{27}\,\mathrm{g}$ as \citet{gronow2020a} for
the base resolution. For each model the highest level of refinement is placed
around the negative $z$-axis at $4\times10^8\,\mathrm{cm}$. The mass resolution
in this region is included in Tables~\ref{tab:model0809} and \ref{tab:model10}
at time $t=1\,\mathrm{s}$ after He ignition.

\subsection{Postprocessing}
Following the explosion simulation, a postprocessing step is carried out. The
evolution of temperature and density in the exploding WD is recorded by two
million tracer particles with a mass of about $1\times10^{27}\,\mathrm{g}$
each, that are randomly distributed in the initial WD
\citep[see][]{travaglio2004a}. The tracer particles are used to determine the
final yields and the chemical structure of the ejecta for subsequent radiative
transfer calculations.

For the postprocessing step a nuclear reaction network with 384 nuclear species
is used. To achieve a more accurate treatment of the metallicity in the WD, all
elements heavier than fluorine are included as given in \citet{asplund2009a}.
We use the 2014 version of the REACLIB data base \citep{rauscher2000a} to
include all relevant reaction rates. This is done using the same method as
\citet{pakmor2012b}.

\section{Models}
\label{sec:models}

\subsection{Model setup}
The study in this paper comprises thirteen models. They cover different shell
and core masses and range from $0.8$\,$M_\odot$ to $1.1$\,$M_\odot$ for the
core and from $0.02$\,$M_\odot$ to $0.1$\,$M_\odot$ for the shell mass of the
WD. The core mass limits correspond to low as well as high luminosity models
\citep[see][for comparison]{sim2010a,fink2010a}. We further cover the highest
expected He shell masses \citep{woosley2011b,neunteufel2016a}, but also reach
down to low shell masses \citep[e.g.][]{fink2010a,townsley2019a}. The
core-shell mass ratio is a parameter that is not well constrained. It highly
depends on the progenitor evolution and ignition process.

The initial models were created in the same way as described in
\citet{gronow2020a}. They were set up to be in hydrostatic equilibrium in 1D.
For this, the total mass ($M_\text{tot}$) and the density ($\rho_\text{s}$)
marking the transition between core and shell are chosen. These in turn
determine the mass of the core ($M_\text{c}$) and shell ($M_\text{s}$) as well
as central density ($\rho_\text{c}$). The core temperature $T_\mathrm{c}$ is
set to a constant value of $3\times10^7\,\mathrm{K}$. In the transition region
between core and shell which spans over $1.8\times10^6\,\mathrm{cm}$ the
temperature changes linearly along with the composition. The temperature at the
base of the He shell $T_\mathrm{s}$ is set to $6\times10^7\,\mathrm{K}$. Beyond
this point, the temperature declines adiabatically. The 1D profile is mapped
to the 3D computational grid of \textsc{Arepo} using the HEALPix method
\citep{gorski2005a} following the procedure of \citet{ohlmann2017a}.  The
metallicity-dependent effect of the electron fraction $Y_e$ is implemented in
this mapping step by adding $^{22}$Ne to the composition of the core material.
This slight perturbation of the hydrostatic equilibrium is leveled off in the
subsequent relaxation step (see Section~\ref{sec:relaxation}).

The parameters of the pre-explosion models are given in
Tables~\ref{tab:model0809} and \ref{tab:model10}. These include the total mass
of the WD ($M_\text{tot}$), the initial core mass ($M_\text{c\_ini}$) and shell
mass ($M_\text{s\_ini}$) after the mapping to the 3D hydrodynamic grid. Grid
cells with an initial He mass fraction of at least 0.01 are considered to be
part of the shell. For completeness the core temperature and density, and
temperature and density at the base of the He shell are listed. The model names
include the initial core and shell mass in the first two and last two digits,
respectively. A WD with an $0.8$\,$M_\odot$ core and $0.03$\,$M_\odot$ shell
initially is therefore named M08\_03.

\begin{table*}[h] 
    \caption{Overview of parameters for models with core masses of about
    $0.8$\,$M_\odot$ and $0.9$\,$M_\odot$.}
\label{tab:model0809}
\centering
\begin{tabular}{@{}lr|rrrr|rrrr}
\hline
& & M08\_10\_r & M08\_10 & M08\_05 & M08\_03 & M09\_10\_r & M09\_10 & M09\_05 & M09\_03\\ \hline
$M_{\mathrm{c\_ini}}$ & [$M_\odot$] & $0.795$ & $0.795$ & $0.803$ & $0.803$ & $0.888$ & $0.888$ & $0.899$ & $0.905$ \\
$M_{\mathrm{s\_ini}}$ & [$M_\odot$] & $0.109$ & $0.109$ & $0.053$ & $0.028$ & $0.108$ & $0.108$ & $0.053$ & $0.026$ \\
$M_{\mathrm{s\_det}}$ & [$M_\odot$] & $0.109$ & $0.127$ & $0.075$ & $0.040$ & $0.108$ & $0.142$ & $0.074$ & $0.043$ \\
$M_{\mathrm{tot}}$ & [$M_\odot$] & $0.910$ & $0.910$ & $0.856$ & $0.830$ & $1.001$ & $1.001$ & $0.952$ & $0.931$ \\
$T_\mathrm{s}$ & [$10^7\,\mathrm{K}$] & $6$ & $6$ & $6$ & $6$ & $6$ & $6$ & $6$ & $6$ \\
$T_\mathrm{c}$ & [$10^7\,\mathrm{K}$] & $3$ & $3$ & $3$ & $3$ & $3$ & $3$ & $3$ & $3$ \\
$\rho_\mathrm{c}$ & [$10^7$\,$\text{g cm}^{-3}$] & $1.864$ & $1.887$ & $1.413$ & $1.224$ & $3.219$ & $3.273$ & $2.471$ & $2.170$ \\
$\rho_\mathrm{s}$ & [$10^6$\,$\text{g cm}^{-3}$] & $0.730$ & $1.034$ & $0.390$ & $0.356$ & $1.303$ & $2.261$ & $0.781$ & $0.493$ \\
$r_\mathrm{det}$ & [$10^8$\,$\text{cm}$] & $4.40$ & $4.48$ & $5.32$ & $5.56$ & $4.21$ & $4.21$ & $4.59$ & $5.02$ \\
He det ign vol & [$10^{23}\,\mathrm{cm}^3$] & $0.12$ & $0.43$& $1.52$ & $6.36$ & $0.40$ & $0.26$ & $1.32$ & $1.26$ \\
$M(^4$He$_{\mathrm{det\_s}})$ & [$M_\odot$] & $0.083$ & $0.082$ & $0.051$ & $0.027$ & $0.085$ & $0.085$ & $0.053$ & $0.026$ \\
$M(^{12}$C$_{\mathrm{det\_s}})$ & [$M_\odot$] & $0.013$ & $0.023$ & $0.012$ & $0.006$ & $0.012$ & $0.029$ & $0.011$ & $0.009$ \\
$M(^{14}$N$_{\mathrm{det\_s}})$ & [$M_\odot$] & $2.7\mathrm{e}{-4}$ & $2.6\mathrm{e}{-4}$ & $1.6\mathrm{e}{-4}$ & $7.7\mathrm{e}{-5}$ & $2.8\mathrm{e}{-4}$ & $2.8\mathrm{e}{-4}$ & $1.7\mathrm{e}{-4}$ & $7.1\mathrm{e}{-5}$ \\
$M(^{16}$O$_{\mathrm{det\_s}})$ & [$M_\odot$] & $0.012$ & $0.022$ & $0.012$ & $0.006$ & $0.011$ & $0.028$ & $0.010$ & $0.009$ \\
$M(^{22}$Ne$_{\mathrm{det\_s}})$ & [$M_\odot$] & $3.4\mathrm{e}{-4}$ & $6.1\mathrm{e}{-4}$ & $3.2\mathrm{e}{-4}$ & $1.7\mathrm{e}{-4}$ & $3.1\mathrm{e}{-4}$ & $7.7\mathrm{e}{-4}$ & $2.8\mathrm{e}{-4}$ & $2.3\mathrm{e}{-4}$ \\
$M(^{12}$C$_{\mathrm{det\_c}})$ & [$M_\odot$] & $0.401$ & $0.392$& $0.393$ & $0.398$ & $0.446$ & $0.429$ & $0.442$ & $0.446$ \\
$M(^{16}$O$_{\mathrm{det\_c}})$ & [$M_\odot$] & $0.390$ & $0.381$& $0.383$ & $0.387$ & $0.434$ & $0.418$ & $0.430$ & $0.434$ \\
$M(^{22}$Ne$_{\mathrm{det\_c}})$ & [$M_\odot$] & $0.011$ & $0.010$ & $0.011$ & $0.011$ & $0.012$ & $0.011$ & $0.012$ & $0.012$ \\
resolution & [$10^{-8}$\,$M_\odot$] & $1.47$ & $2.08$ & $3.95$ & $37.10$ & $1.38$ & $4.89$ & $2.26$ & $4.34$ \\
ignition mechn. & & s & & (s,) cs & cs & s & & (s,) cs & (s,) cs \\
core ign. time & & $1.33$ & $1.102$ & $2.05$ & $2.65$ & $1.17$ & $0.50$ & $1.71$ & $2.14$ \\
\hline
\end{tabular}
\end{table*}

\begin{table*}[h]
    \caption{Overview of parameters for models with a core mass of about $1.0$\,$M_\odot$ and $1.1$\,$M_\odot$.}
    \label{tab:model10}
    \centering
    \begin{tabular}{@{}lr|rrrrr}
        \hline
        & & M11\_05 & M10\_10 & M10\_05 & M10\_03 & M10\_02 \\ \hline
        $M_{\mathrm{c\_ini}}$ & [$M_\odot$] & $1.100$ & $1.015$ & $1.002$ & $1.028$ & $1.005$ \\
        $M_{\mathrm{s\_ini}}$ & [$M_\odot$] & $0.054$ & $0.090$ & $0.052$ & $0.027$ & $0.020$ \\
        $M_{\mathrm{s\_det}}$ & [$M_\odot$] & $0.123$ & $0.133$ & $0.074$ & $0.047$ & $0.028$ \\
        $M_{\mathrm{tot}}$ & [$M_\odot$] & $1.159$ & $1.105$ & $1.055$ & $1.055$ & $1.025$ \\
        $T_\mathrm{s}$ & [$10^7\,\mathrm{K}$] & $6$ & $6$ & $6$ & $6$ & $6$ \\
        $T_\mathrm{c}$ & [$10^7\,\mathrm{K}$] & $3$ & $3$ & $3$ & $3$ & $3$ \\
        $\rho_\mathrm{c}$ & [$10^7$\,$\text{g cm}^{-3}$] & $10.213$ & $6.847$ & $4.777$ & $4.777$ & $3.904$ \\
        $\rho_\mathrm{s}$ & [$10^6$\,$\text{g cm}^{-3}$] & $2.000$ & $2.460$ & $1.094$ & $0.850$ & $0.510$ \\
        $r_\mathrm{det}$ & [$10^8$\,$\text{cm}$] & $3.53$ & $3.47$ & $4.20$ & $4.25$ & $4.36$ \\
        He det ign vol & [$10^{23}\,\mathrm{cm}^3$] & $0.22$ & $0.63$ & $0.15$ & $0.77$ & $0.95$ \\
        $M(^4$He$_{\mathrm{det\_s}})$ & [$M_\odot$] & $0.049$ & $0.084$ & $0.050$ & $0.026$ & $0.020$ \\
        $M(^{12}$C$_{\mathrm{det\_s}})$ & [$M_\odot$] & $0.037$ & $0.024$ & $0.012$ & $0.010$ & $0.004$ \\
        $M(^{14}$N$_{\mathrm{det\_s}})$ & [$M_\odot$] & $1.5\mathrm{e}{-4}$ & $2.7\mathrm{e}{-4}$ & $1.5\mathrm{e}{-4}$ & $7.3\mathrm{e}{-5}$ & $5.2\mathrm{e}{-5}$ \\
        $M(^{16}$O$_{\mathrm{det\_s}})$ & [$M_\odot$] & $0.036$ & $0.023$ & $0.012$ & $0.010$ & $0.004$ \\
        $M(^{22}$Ne$_{\mathrm{det\_s}})$ & [$M_\odot$] & $0.001$ & $6.4\mathrm{e}{-4}$ & $3.3\mathrm{e}{-4}$ & $2.8\mathrm{e}{-4}$ & $1.1\mathrm{e}{-4}$ \\
        $M(^{12}$C$_{\mathrm{det\_c}})$ & [$M_\odot$] & $0.518$ & $0.489$ & $0.493$ & $0.506$ & $0.501$ \\
        $M(^{16}$O$_{\mathrm{det\_c}})$ & [$M_\odot$] & $0.504$ & $0.475$ & $0.479$ & $0.493$ & $0.487$ \\
        $M(^{22}$Ne$_{\mathrm{det\_c}})$ & [$M_\odot$] & $0.014$ & $0.013$ & $0.013$ & $0.014$ & $0.013$ \\
        resolution & [$10^{-8}$\,$M_\odot$] & $27.36$ & $78.11$ & $3.38$ & $3.61$ & $47.71$ \\
        ignition mechn. & & edge & edge & s & (s,) cs & art cs \\
        core ign. time & & 0.006 & $0.005$ & $1.17$ & $1.62$ & $1.96$ \\
    \end{tabular}
\end{table*}

\subsection{Relaxation and treatment of core~-~shell mixing}
\label{sec:relaxation}
\begin{figure}
\centering
   \includegraphics[width=0.45\textwidth]{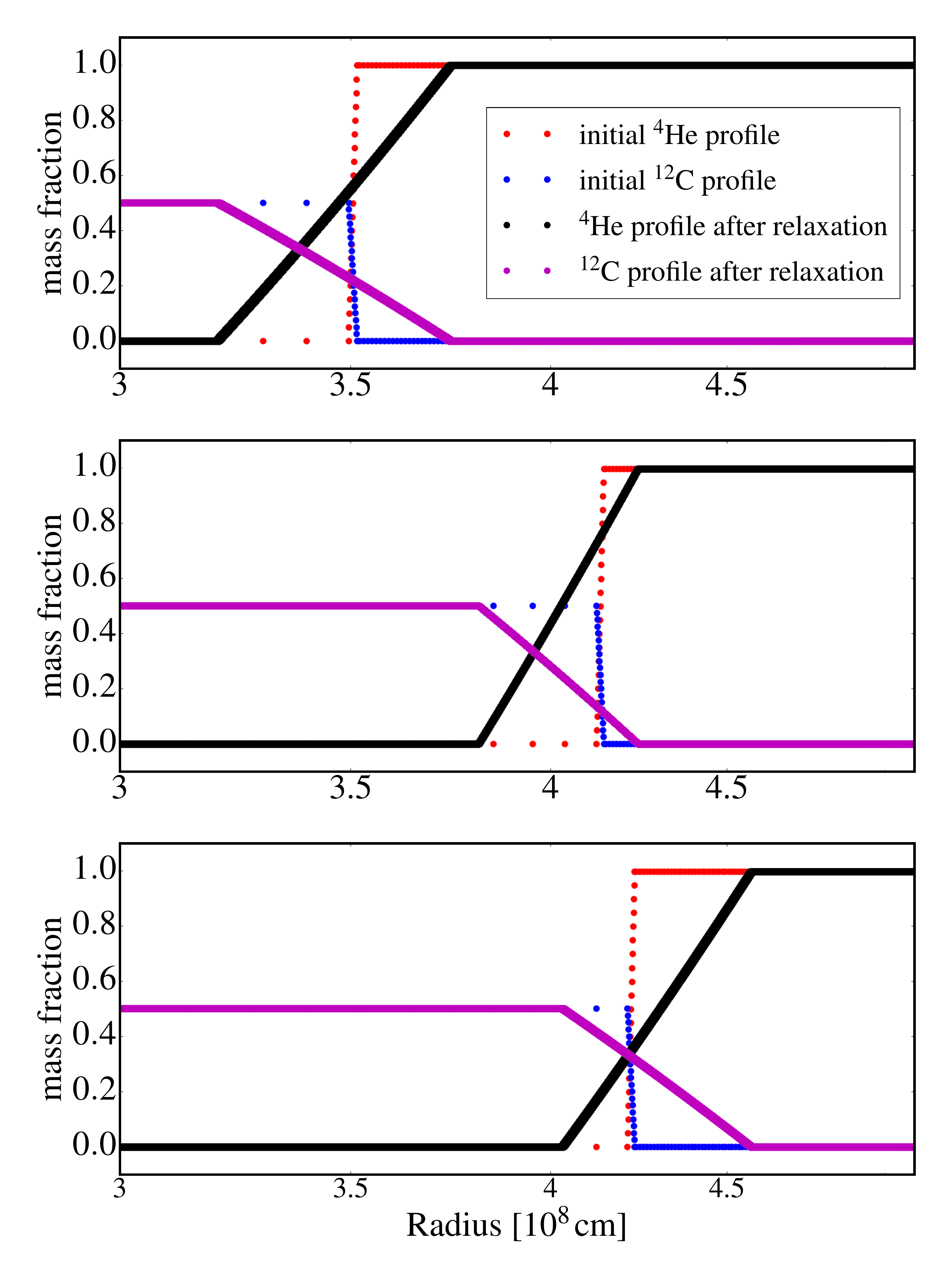}
   \caption{The radial abundance profiles of $^4$He and $^{12}$C for Model
   M10\_10, M10\_05 and M10\_03 from top to bottom; the initial profiles before
   relaxation are shown in red and blue, and the profiles after relaxation in
   black and magenta.}
      \label{fig:profiles}
\end{figure}

To remove spurious velocities that might occur due to the mapping of the 1D
setup onto the 3D computational grid, a relaxation step according to
\citet{ohlmann2017a} is carried out, as also implemented in
\citet{gronow2020a}.  Spurious velocities are damped by the addition of a
source term proportional to
\begin{align*}
    \dot{\vec{v}} = -\frac{1}{\tau}\vec{v}
\end{align*}
to the momentum equation as stated in Equation (8) of \citet{ohlmann2017a} with
the damping timescale $\tau$. The degree of damping is decreased during the
relaxation step until eight dynamical timescales have passed. No damping is
applied during the remaining relaxation time until ten dynamical time scales
have passed. This allows us to check for each model whether the stability
criteria listed in \citet{ohlmann2017a} are met. The relaxation leads to
additional mixing between the core and shell. We find that the amount of mixing
primarily depends on the initial shell mass as the models with a similar
initial shell mass show a good agreement in the shell composition after the
relaxation.  Because all grid cells with a helium mass fraction of at least
$0.01$ are considered to be part of the shell, the shell formally increases in
mass during this step. The abundance profiles of $^{4}$He and $^{12}$C are
shown in Figure~\ref{fig:profiles} for Models M10\_10, M10\_05, and M10\_03
(see Table~\ref{tab:model10} for details).  The profiles after relaxation (in
black and magenta) have a much broader transition region between core and shell
material. It is also visible that the base of the shell has moved further in
during the relaxation. The compositions of shell and core after the relaxation
are given in Tables~\ref{tab:model0809} and \ref{tab:model10} by
$M(^4$He$_{\mathrm{det\_s}})$, $M(^{12}$C$_{\mathrm{det\_s}})$,
$M(^{14}$N$_{\mathrm{det\_s}})$, $M(^{16}$O$_{\mathrm{det\_s}})$ and
$M(^{22}$Ne$_{\mathrm{det\_s}})$, and $M(^{12}$C$_{\mathrm{det\_c}})$,
$M(^{16}$O$_{\mathrm{det\_c}})$ and $M(^{22}$Ne$_{\mathrm{det\_c}})$,
respectively. In our models, the degree of mixing between core and shell is set
by the relaxation step, in addition to a small initial transition region in the
1D profile. In reality, this mixing depends on complex processes in the
accretion phase and its strength is not known to date.  Simulations of rotating
WDs by \citet{neunteufel2017a} consider aspects such as dynamical shear
instability, Goldreich-Schubert-Fricke instability and secular shear
instability to model mixing appropriately \citep[see][for
details]{neunteufel2017a}. The simulations show that the degree of mixing
depends on many parameters. The strongest effect is seen with a change in total
mass (less massive systems showing more mixing than massive systems).  Due to
the uncertainties of the exact mixing the setup after relaxation is used as a
first test.

The effect of core-shell mixing is analysed in more detail with the
consideration of two additional models,  M08\_10\_r  and M09\_10\_r, for which
the composition was reset. After relaxing models M08\_10 and M09\_10, the
composition is changed back to the initial profiles of the corresponding 1D
setups (see Table~\ref{tab:model0809} for the values). By comparing these
de-mixed models to their counterparts in our standard setup, M08\_10 and
M09\_10, we can assess the impact of the assumptions made for explosion
dynamics and the derived nucleosynthetic yields.  Note that the model
parameters differ slightly in the standard and de-mixed setups: due to the
relaxation procedure described above some core material is mixed into the shell
causing the shell mass to increase and the core radius (defined as the maximum
radius with a He mass fraction lower than $0.01$) to decrease.

\subsection{Detonation simulations}
\label{sec:detonation}
At the beginning of the explosion simulation the He detonation is ignited
artificially in one roughly spherical region around the radius of the peak in
the temperature profile. The volume of the He detonation ignition region is
included in Tables~\ref{tab:model0809} and \ref{tab:model10}, and is set to
have a previously set value. The He detonation is ignited by increasing the
specific thermal energy around the temperature peak as explained in
\citet{gronow2020a} and the location of the ignition spot is chosen to be on
the positive $z$-axis with $x=y=0$. The radial position of the center of the He
detonation region is given in Tables~\ref{tab:model0809} and \ref{tab:model10}.
The evolution is followed for $100\,\mathrm{s}$.

\section{Results}
\label{sec:results}
Our parameter study covers a range of WD core and shell masses. In the
following we discuss the effect on the detonation ignition mechanism and final
abundances.

\subsection{Detonation ignition mechanism}
\label{sec:IgnMech}
\begin{figure*}
    \centering
    \includegraphics[width=0.96\textwidth]{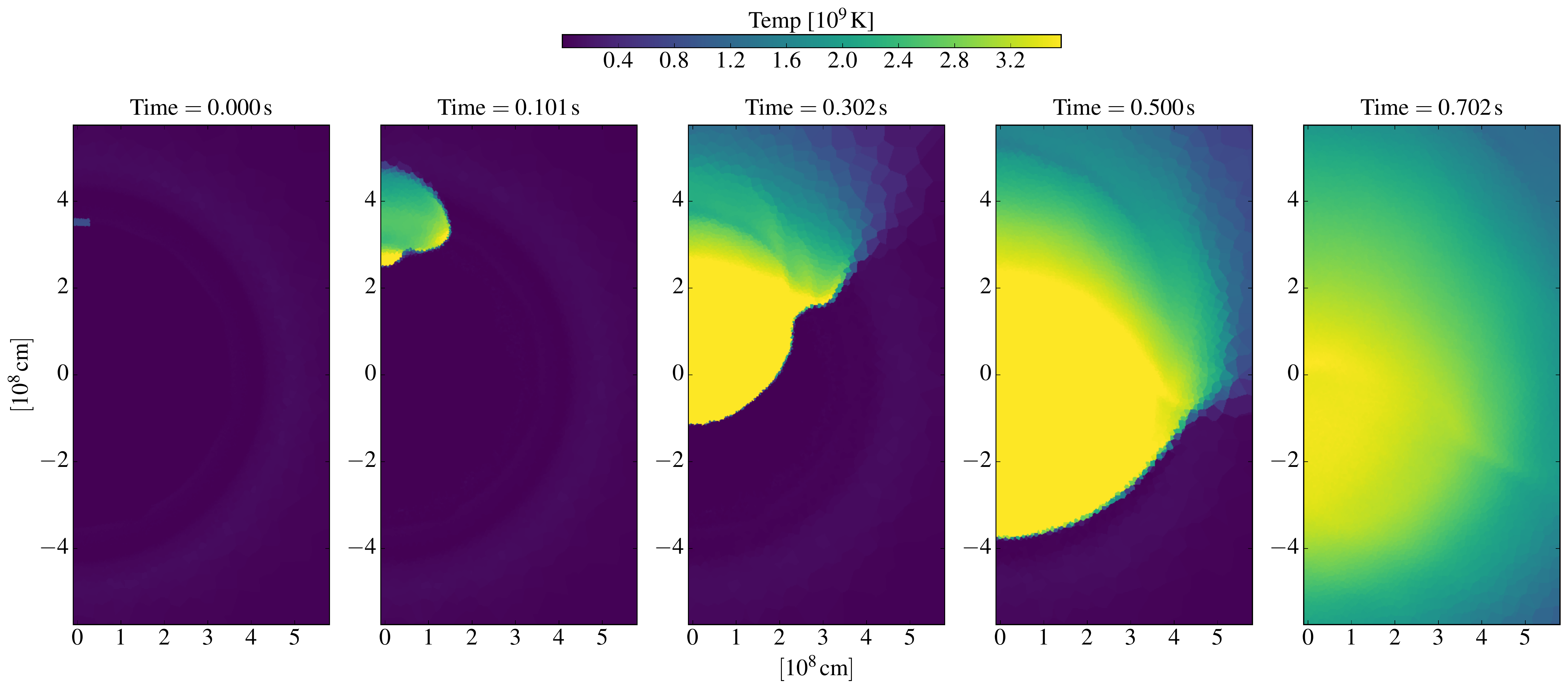}

    \includegraphics[width=0.96\textwidth]{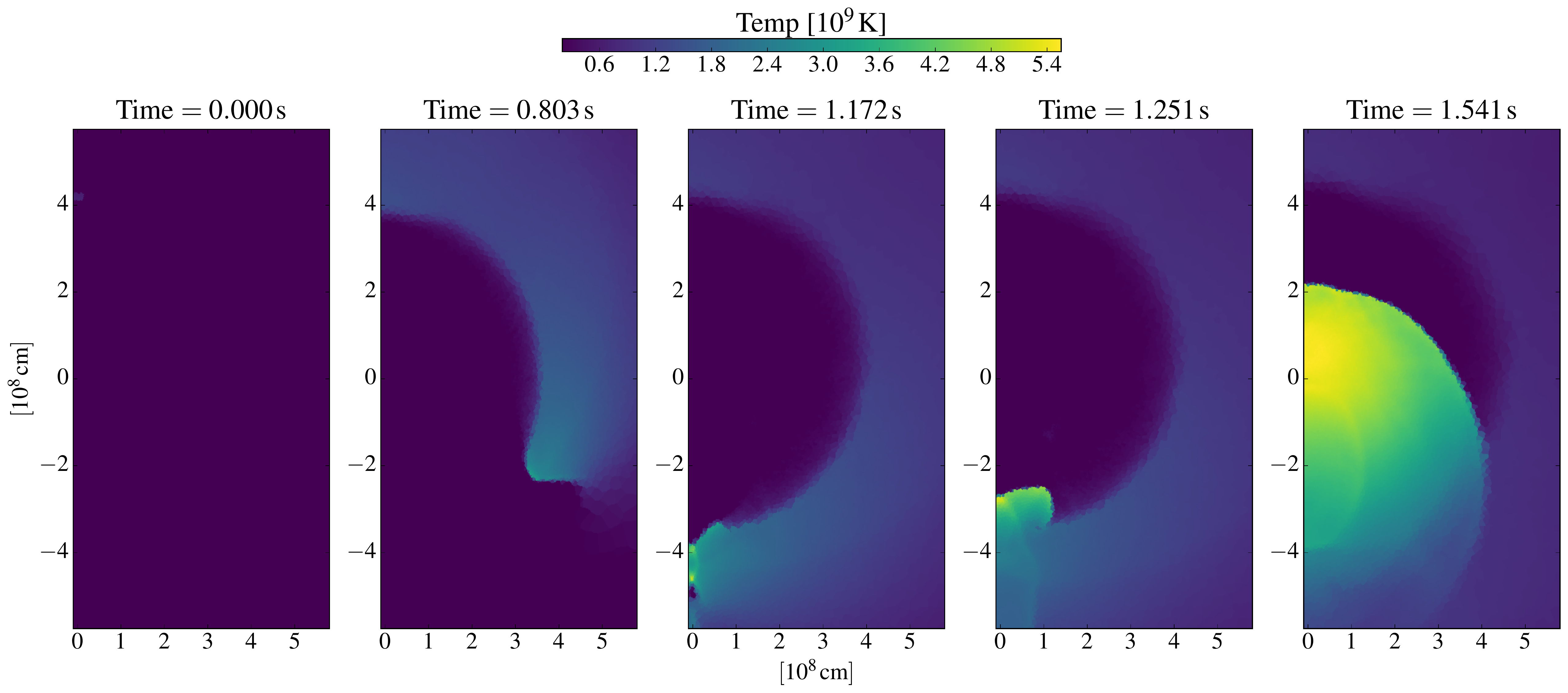}

    \includegraphics[width=0.96\textwidth]{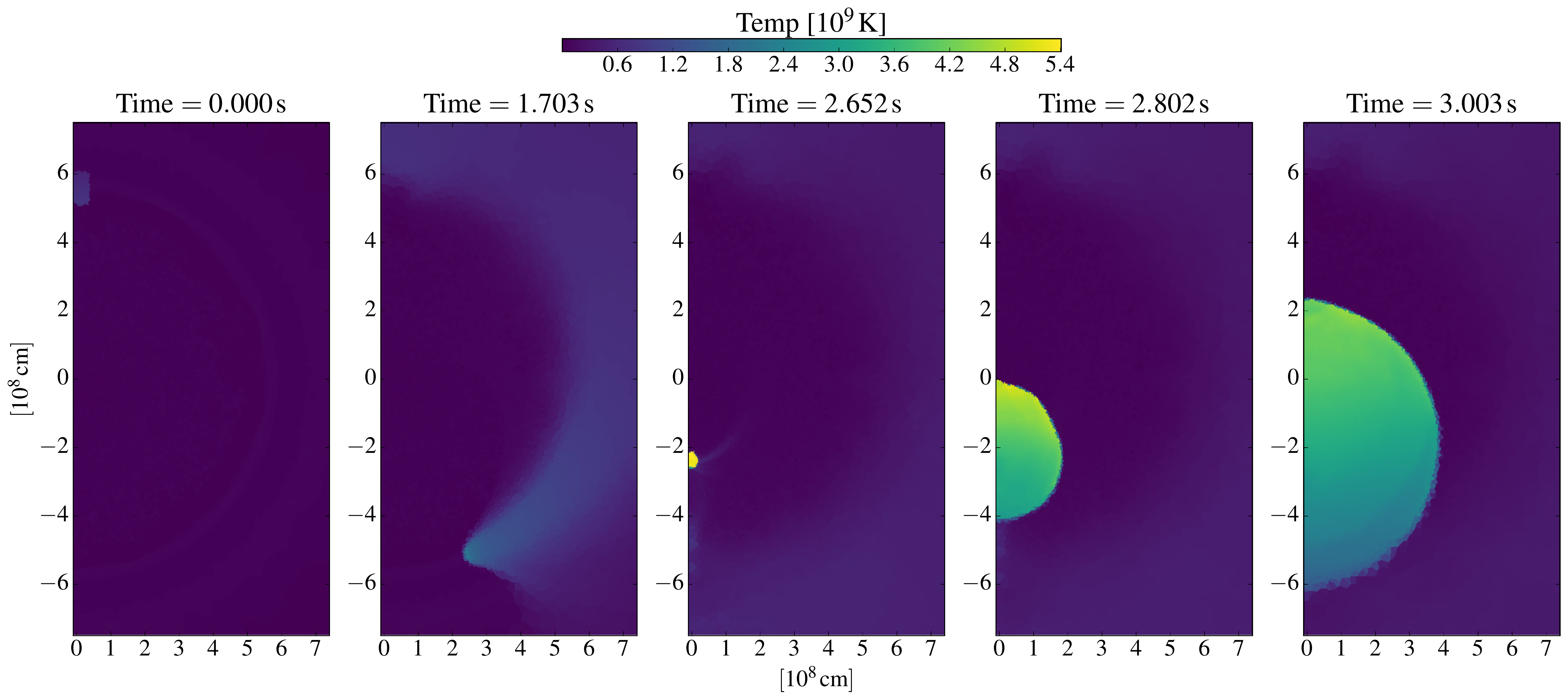}
    \caption{Time evolution of Model M10\_10, M10\_05 and M08\_03 from top to
        bottom showing the edge-lit, scissors and converging shock mechanism,
        respectively; the temperature is given in K at different times
        increasing from left to right in a cut through the center of the WD
        showing only the positive $x$ axis.}
    \label{fig:mechanisms}
\end{figure*}

The double detonation scenario consists of two detonations. The first
detonation is in the He shell. \citet{shen2010a} and \citet{glasner2018a}
describe a mechanism leading to He ignition: The shell material is heated by
compression due to the accretion process and convection sets in. The material
is unstable to convection, and temperature fluctuations develop because of
convective burning. The He burning increases the temperature further,
increasing the burning rates. In hotspots this results in a burning time scale
shorter than the convective turn over time and dynamical time scale, and allows
a detonation to develop. \citet{shen2014b} investigate different sizes for
hotspots as well as the effect of possible pollution of the material by carbon
and oxygen.

Following its ignition the He detonation propagates in the shell around the
core driving a shock wave into the core. The He detonation converges on the far
side of its ignition spot and a shock wave moves into the core. The shock waves
converge off-center in the WD core at high densities of about
$2.0\times10^7\,\text{g cm}^{-3}$. At this point a core detonation is ignited
which then burns through the whole star. This detonation mechanism is called
the `converging shock scenario' (`cs' in the tables)
\citep[e.g.][]{livne1990a,fink2007a,shen2014a}.

However, there are a number of mechanisms that can ignite the core and
have previously been discussed in the literature. In the `edge-lit scenario'
(`edge') \citep[e.g.][]{livne1990b} a core detonation is ignited at the
core-shell interface directly after the ignition of the He detonation. A third
mechanism is the `scissors scenario' (`s') \citep{forcada2006a,gronow2020a}.
In this case the convergence of the He detonation wave takes place in a
carbon-enriched transition region between core and shell. It is strong enough
to ignite carbon burning at this point, before the convergence of the shock
waves in the core. We observe all three detonation ignition mechanisms. The
mechanism of each model is listed in Tables~\ref{tab:model0809} and
\ref{tab:model10}.

The exact ignition mechanism of the core detonation depends on the specific
setup of the WD. For example, the density at the base of the He shell is an
important parameter for the edge-lit mechanism. The amount of carbon mixed into
the shell and the details of the transition region between core and shell are
also important, in particular for the scissors mechanism. The carbon detonation
ignition is not fully resolved in the current simulations and the observed core
ignition is at least in parts a numerical artifact. \citet{katz2019a} argue
that a resolution of 1\,km is needed.  \citet{kushnir2020a}, in contrast, state
that a 10\,km resolution is sufficient when using their burning limiter. The
cells showing a carbon ignition in our simulations have a radial extent of some
30\,km assuming a sphere in our models.  Therefore, we check whether sensible
values for an ignition \citep{roepke2007a,seitenzahl2009b} are reached. It is
reasonable to trust the different mechanisms found in simulations if critical
values are reached in large enough regions. With this additional constraint the
core detonations can be interpreted as being physical, although this is not
rigorously proven by our simulations.

A second, core detonation is observed in all models except for the model with
the lowest He shell mass of initially $0.02$\,$M_\odot$, M10\_02, where a core
detonation is not triggered. In this case, a core detonation is ignited
artificially when the density and temperature reach values of at least
$2.5\times10^7$\,$\text{g cm}^{-3}$ and $8.0\times10^8\,\mathrm{K}$,
respectively. \citet{townsley2019a} have claimed a successful core detonation
using a similar setup. We do not confirm this in our simulation, which can be
caused by a lack of resolution. But in order to test the potential outcome in
the case of a core detonation ignition, a carbon detonation is ignited once
critical values found in \citet{roepke2007a} and \citet{seitenzahl2009b} are
met. Since these values are reached in some cells, this indicates that a
detonation ignition may be physical in this case. It shows that the artificial
numerical ignition fails for this model. The ignition mechanism of the core
detonation found for each model is listed in Tables~\ref{tab:model0809} and
\ref{tab:model10}, as well as the time of carbon ignition.

The edge-lit mechanism is found in the models with the highest masses, both for
core and shell, M10\_10 and M11\_05. The propagation of the double detonation
in Model M10\_10 can be inferred from the evolution of the temperature as
illustrated in the top row of Figure~\ref{fig:mechanisms}. A comparison of the
two leftmost panels shows that the carbon detonation is ignited shortly after
the He detonation. By the time of the rightmost panel ($0.7\, \mathrm{s}$ after
He detonation ignition), the detonation has propagated through the whole core.
The He detonation was ignited at the peak of the initial temperature profile.
With $2.2\times10^6\,\text{g cm}^{-3}$, the density at this point is the
second highest of all models in our study (with only M11\_05 reaching a higher
value). Since the temperature is increased to values of at least
$7.0\times10^8\,\mathrm{K}$, it is sufficient to cause a carbon detonation in
the transition region between core and shell. \citet{livne1990b} argue that the
He detonation needs to be ignited at some altitude above the core-shell
interface as the detonation needs some time to develop enough strength to
ignite the core. In contrast we find that the He detonation fades out if it is
ignited further out. This is due to the lower density. However, the ignition
spot is located $2.7\times10^7\,\mathrm{cm}$ away from the base of the shell
allowing the detonation to be strong enough for a carbon ignition.

The carbon detonation in Models M08\_10\_r, M09\_10\_r and M10\_05 is ignited
via the scissors mechanism \citep[see][for a detailed
description]{gronow2020a}. M10\_05 is set up in the same way as Model M2a of
\citet{gronow2020a}. The only difference here is the solar metallicity of the
stellar material in our new model. This leads to slight changes in the
composition of the shell after the relaxation compared to Model M2a. These
differences are, however, too small to affect or alter the ignition mechanism.
After the He detonation is ignited (left panel of Figure~\ref{fig:mechanisms},
second row) and propagates through the shell (second panel from left), carbon
is ignited (central panel) at the interface between core and shell. Here the
carbon mass fraction is at least 0.29 at a point with a density of
$5.0\times10^6$\,$\text{g cm}^{-3}$ and temperature of
$2.5\times10^9\,\mathrm{K}$ which fulfills the detonation ignition conditions
put forward in \citet{roepke2007a} and \citet{seitenzahl2009b}.

The effect of mixing between core and shell on the carbon ignition mechanism is
studied with the comparison of Models M08\_10 and M08\_10\_r as well as M09\_10
and M09\_10\_r (see Section~\ref{sec:relaxation} for details). The total (and
shell) masses of M08\_10 and M09\_10 are among the highest in our parameter
study. While the composition of the shell is comparable to M10\_10, the He
detonation burns at a much lower density which does not allow a direct ignition
of the core as in M10\_10. The models, instead, illustrate the importance of
the details of the transition region and the degree of mixing between core and
shell: It is observed that the He detonation burns at the very base of the
shell. Because the composition changes over a large radial span in the
transition of M08\_10, it reaches regions with high enough $^{12}$C mass
fraction and densities to ignite carbon already when the He detonation only
propagated around two thirds of the WD. Following this, a carbon detonation
moves into the core.  In order to see whether this kind of carbon ignition can
be physical, a higher resolution study of the transition region is required.
This study goes beyond the scope of this work and will be carried out in the
future. Furthermore, it is necessary to consider different sizes of the
transition region, and with that different degrees of mixing, as the slope of
the change in carbon abundance is an important parameter. A shallower change
makes the propagation of the He detonation into higher density carbon enriched
material easier.

Our results confirm that the details of the transition region are an important
parameter for the carbon ignition as already stated in \citet{gronow2020a}.
M08\_10\_r has a thinner shell than M08\_10 containing less carbon and
corresponding to that a smaller transition region. Additionally, the density at
the base of the He shell is lower, prohibiting a carbon detonation early at
the interface. Instead, carbon is ignited by the scissors mechanism. The same
behaviour as found for M08\_10\_r can be observed in M09\_10\_r.

Four models, M08\_05, M09\_05, M09\_03 and M10\_03, show some burning of
carbon producing heavy elements in the region where the He detonation wave
converges at the antipode of the He detonation ignition spot. This is similar
to the scissors mechanism and marked by `(s,)' in Tables~\ref{tab:model0809}
and \ref{tab:model10}. However, the burning is not strong enough for a core
detonation to start. Instead, the shock wave converges off-center in the core
before the convergence of the He detonation wave can ignite a successful carbon
detonation. This core detonation follows the converging shock mechanism.
Unlike M10\_05 and M2a, the models have lower densities at the core-shell
interface. The conditions listed in \citet{seitenzahl2009b} are therefore not
reached for a successful carbon ignition. Only two cells match the conditions
in M10\_03. This is however not sufficient for an ignition in our numerical
treatment. It is to be noted that we refer to the lowest critical values found
in \citet{seitenzahl2009b} which indicates that they are too low for a carbon
ignition in the setup of M10\_03. The burning front propagates much slower in
these models.

Model M08\_03 does not show this behavior: The He detonation burns at the
lowest density compared to the other models of the parameter study so that
conditions to ignite carbon in the core--shell interface are not met at the
antipode of the He ignition spot. Only the convergence of the shock wave in the
core is strong enough to reach critical values and ignite a carbon detonation.
Model M10\_02 tests the limit of an extremely low-mass He shell. Here, all
potential carbon ignition mechanisms fail. Even the central convergence of the
shock wave in Model M10\_02 is not strong enough for an ignition. A core
detonation is rather ignited artificially (marked as `art cs' in Tab.
\ref{tab:model10}) when critical values listed above are reached. In this case,
values are attained for a physical ignition, but are not high enough to trigger
a numerical carbon ignition in the \textsc{Arepo} code. However, it is possible
that an ignition can be observed in a higher resolution simulation.

\subsection{Final abundances}
\label{sec:abundances}

\begin{table*}
    \caption{Final abundances for Model M08\_10, M08\_10\_r, M08\_05 and M08\_03.}
    \label{tab:abund08}
    \centering
    \begin{tabular}{@{}lrrrrrrrr}
        \hline
        & \multicolumn{4}{c}{He detonation} & \multicolumn{4}{c}{core detonation} \\
        & M08\_10 & M08\_10\_r & M08\_05 & M08\_03 & M08\_10 & M08\_10\_r & M08\_05 & M08\_03 \\
        & [$M_\odot$] & [$M_\odot$] & [$M_\odot$] & [$M_\odot$] & [$M_\odot$] & [$M_\odot$] & [$M_\odot$] & [$M_\odot$] \\ \hline
        $^4$He & $3.1\times10^{-2}$ & $3.6\times10^{-2}$ & $2.7\times10^{-2}$ & $1.8\times10^{-2}$ & $1.9\times10^{-3}$ & $1.4\times10^{-3}$ & $8.2\times10^{-5}$ & $3.2\times10^{-6}$\\
        $^{12}$C & $8.5\times10^{-5}$ & $1.2\times10^{-4}$ & $2.3\times10^{-3}$ & $3.3\times10^{-3}$ & $3.0\times10^{-4}$ & $1.1\times10^{-3}$ & $7.5\times10^{-3}$ & $1.2\times10^{-2}$\\
        $^{16}$O & $1.7\times10^{-2}$ & $9.3\times10^{-3}$ & $6.3\times10^{-3}$ & $2.6\times10^{-3}$ & $7.9\times10^{-2}$ & $8.1\times10^{-2}$ & $1.2\times10^{-1}$ & $1.4\times10^{-1}$\\
        $^{28}$Si & $2.3\times10^{-2}$ & $1.3\times10^{-2}$ & $9.3\times10^{-3}$ & $4.2\times10^{-3}$ & $1.9\times10^{-1}$ & $1.9\times10^{-1}$ & $2.3\times10^{-1}$ & $2.6\times10^{-1}$\\
        $^{32}$S & $9.1\times10^{-3}$ & $5.5\times10^{-3}$ & $4.7\times10^{-3}$ & $2.4\times10^{-3}$ & $1.1\times10^{-1}$ & $1.1\times10^{-1}$ & $1.3\times10^{-1}$ & $1.4\times10^{-1}$\\
        $^{40}$Ca & $8.1\times10^{-3}$ & $6.2\times10^{-3}$ & $8.0\times10^{-3}$ & $3.1\times10^{-3}$ & $1.6\times10^{-2}$ & $1.7\times10^{-2}$ & $1.9\times10^{-2}$ & $1.9\times10^{-2}$\\
        $^{44}$Ti & $1.9\times10^{-3}$ & $1.8\times10^{-3}$ & $2.7\times10^{-3}$ & $2.2\times10^{-4}$ & $1.4\times10^{-5}$ & $1.4\times10^{-5}$ & $1.2\times10^{-5}$ & $1.2\times10^{-5}$\\
        $^{48}$Cr & $4.5\times10^{-3}$ & $3.8\times10^{-3}$ & $2.6\times10^{-3}$ & $7.2\times10^{-6}$ & $3.0\times10^{-4}$ & $3.3\times10^{-4}$ & $3.1\times10^{-4}$ & $2.9\times10^{-4}$ \\
        $^{52}$Fe & $8.1\times10^{-3}$ & $7.5\times10^{-3}$ & $8.0\times10^{-4}$ & $8.8\times10^{-7}$ & $6.5\times10^{-3}$ & $7.3\times10^{-3}$ & $6.8\times10^{-3}$ & $5.6\times10^{-3}$ \\
        $^{55}$Mn & $6.5\times10^{-8}$ & $6.4\times10^{-8}$ & $1.0\times10^{-7}$ & $1.8\times10^{-7}$ & $7.9\times10^{-8}$ & $6.8\times10^{-8}$ & $9.9\times10^{-8}$ & $2.0\times10^{-7}$ \\
        $^{55}$Co & $8.7\times10^{-4}$ & $9.4\times10^{-4}$ & $3.1\times10^{-5}$ & $2.8\times10^{-7}$ & $3.9\times10^{-3}$ & $3.8\times10^{-3}$ & $3.5\times10^{-3}$ & $2.8\times10^{-3}$ \\
        $^{56}$Ni & $1.1\times10^{-2}$ & $1.5\times10^{-2}$ & $6.7\times10^{-5}$ & $9.9\times10^{-7}$ & $3.0\times10^{-1}$ & $3.1\times10^{-1}$ & $2.0\times10^{-1}$ & $1.3\times10^{-1}$ \\ \hline
    \end{tabular}
\end{table*}

\begin{table*}
    \caption{Final abundances for Model M09\_10, M09\_10\_r, M09\_05 and M09\_03.}
    \label{tab:abund09}
    \centering
    \begin{tabular}{@{}lrrrrrrrr}
        \hline
        & \multicolumn{4}{c}{He detonation} & \multicolumn{4}{c}{core detonation} \\
        & M09\_10 & M09\_10\_r & M09\_05 & M09\_03 & M09\_10 & M09\_10\_r & M09\_05 & M09\_03 \\
        & [$M_\odot$] & [$M_\odot$] & [$M_\odot$] & [$M_\odot$] & [$M_\odot$] & [$M_\odot$] & [$M_\odot$] & [$M_\odot$] \\ \hline
        $^4$He & $2.6\times10^{-2}$ & $3.2\times10^{-2}$ & $2.5\times10^{-2}$ & $1.5\times10^{-2}$ & $3.2\times10^{-3}$ & $3.9\times10^{-3}$ & $1.9\times10^{-3}$ & $5.8\times10^{-4}$\\
        $^{12}$C & $3.1\times10^{-5}$ & $3.9\times10^{-5}$ & $4.3\times10^{-4}$ & $3.5\times10^{-3}$ & $2.0\times10^{-6}$ & $1.3\times10^{-4}$ & $2.6\times10^{-3}$ & $4.9\times10^{-3}$\\
        $^{16}$O & $1.5\times10^{-2}$ & $8.5\times10^{-3}$ & $7.3\times10^{-3}$ & $3.9\times10^{-3}$ & $2.7\times10^{-2}$ & $5.5\times10^{-2}$ & $7.8\times10^{-2}$ & $9.2\times10^{-2}$\\
        $^{28}$Si & $3.9\times10^{-2}$ & $1.3\times10^{-2}$ & $1.0\times10^{-2}$ & $5.8\times10^{-3}$ & $1.5\times10^{-1}$ & $1.6\times10^{-1}$ & $1.9\times10^{-1}$ & $2.2\times10^{-1}$\\
        $^{32}$S & $1.1\times10^{-2}$ & $4.3\times10^{-3}$ & $4.4\times10^{-3}$ & $2.8\times10^{-3}$ & $9.4\times10^{-2}$ & $9.2\times10^{-2}$ & $1.1\times10^{-1}$ & $1.3\times10^{-1}$\\
        $^{40}$Ca & $7.9\times10^{-3}$ & $4.7\times10^{-3}$ & $5.1\times10^{-3}$ & $4.0\times10^{-3}$ & $1.7\times10^{-2}$ & $1.6\times10^{-2}$ & $1.8\times10^{-2}$ & $2.0\times10^{-2}$\\
        $^{44}$Ti & $8.5\times10^{-4}$ & $8.9\times10^{-4}$ & $2.0\times10^{-3}$ & $7.2\times10^{-4}$ & $1.6\times10^{-5}$ & $1.6\times10^{-5}$ & $1.5\times10^{-5}$ & $1.4\times10^{-5}$\\
        $^{48}$Cr & $2.5\times10^{-3}$ & $1.9\times10^{-3}$ & $4.6\times10^{-3}$ & $1.0\times10^{-4}$ & $3.7\times10^{-4}$ & $3.4\times10^{-4}$ & $3.7\times10^{-4}$ & $3.9\times10^{-4}$ \\
        $^{52}$Fe & $5.1\times10^{-3}$ & $4.0\times10^{-3}$ & $5.1\times10^{-3}$ & $4.1\times10^{-6}$ & $8.2\times10^{-3}$ & $7.5\times10^{-3}$ & $8.1\times10^{-3}$ & $8.8\times10^{-3}$ \\
        $^{55}$Mn & $7.6\times10^{-8}$ & $6.2\times10^{-8}$ & $6.8\times10^{-8}$ & $1.3\times10^{-7}$ & $1.8\times10^{-8}$ & $4.5\times10^{-8}$ & $5.7\times10^{-8}$ & $7.6\times10^{-8}$ \\
        $^{55}$Co & $4.5\times10^{-4}$ & $3.7\times10^{-4}$ & $4.1\times10^{-4}$ & $4.3\times10^{-7}$ & $4.9\times10^{-3}$ & $3.9\times10^{-3}$ & $4.2\times10^{-3}$ & $4.5\times10^{-3}$ \\
        $^{56}$Ni & $2.2\times10^{-2}$ & $2.6\times10^{-2}$ & $2.0\times10^{-3}$ & $1.0\times10^{-6}$ & $4.7\times10^{-1}$ & $4.8\times10^{-1}$ & $3.8\times10^{-1}$ & $3.3\times10^{-1}$ \\ \hline
    \end{tabular}
\end{table*}

\begin{table*}
    \caption{Final abundances for Model M10\_10, M10\_05, M10\_03 and M10\_02.}
    \label{tab:abund10}
    \centering
    \begin{tabular}{@{}lrrrrrrrr}
        \hline
        & \multicolumn{4}{c}{He detonation} & \multicolumn{4}{c}{core detonation} \\
        & M10\_10 & M10\_05 & M10\_03 & M10\_02 & M10\_10 & M10\_05 & M10\_03 & M10\_02 \\
        & [$M_\odot$] & [$M_\odot$] & [$M_\odot$] & [$M_\odot$] & [$M_\odot$] & [$M_\odot$] & [$M_\odot$] & [$M_\odot$] \\ \hline
        $^4$He & $2.1\times10^{-2}$ & $2.0\times10^{-2}$ & $1.3\times10^{-2}$ & $1.3\times10^{-2}$ & $6.5\times10^{-3}$ & $4.6\times10^{-3}$ & $5.1\times10^{-3}$ & $3.8\times10^{-3}$ \\
        $^{12}$C & $1.1\times10^{-5}$ & $4.0\times10^{-5}$ & $7.6\times10^{-4}$ & $1.7\times10^{-3}$ & $1.7\times10^{-5}$ & $4.4\times10^{-4}$ & $1.2\times10^{-3}$ & $1.9\times10^{-3}$ \\
        $^{16}$O & $3.1\times10^{-3}$ & $9.3\times10^{-3}$ & $6.8\times10^{-3}$ & $1.9\times10^{-3}$ & $2.7\times10^{-3}$ & $6.1\times10^{-2}$ & $4.9\times10^{-2}$ & $5.7\times10^{-2}$ \\
        $^{28}$Si & $3.7\times10^{-2}$ & $1.3\times10^{-2}$ & $8.9\times10^{-3}$ & $2.9\times10^{-3}$ & $7.3\times10^{-2}$ & $1.6\times10^{-1}$ & $1.5\times10^{-1}$ & $1.7\times10^{-1}$ \\
        $^{32}$S & $1.6\times10^{-2}$ & $4.9\times10^{-3}$ & $3.7\times10^{-3}$ & $1.6\times10^{-3}$ & $5.4\times10^{-2}$ & $9.6\times10^{-2}$ & $9.1\times10^{-2}$ & $1.0\times10^{-1}$ \\
        $^{40}$Ca & $3.4\times10^{-3}$ & $4.3\times10^{-3}$ & $3.3\times10^{-3}$ & $2.4\times10^{-3}$ & $1.3\times10^{-2}$ & $1.7\times10^{-2}$ & $1.6\times10^{-2}$ & $1.8\times10^{-2}$ \\
        $^{44}$Ti & $2.7\times10^{-4}$ & $7.9\times10^{-4}$ & $1.1\times10^{-3}$ & $5.7\times10^{-4}$ & $1.8\times10^{-5}$ & $2.1\times10^{-5}$ & $1.8\times10^{-5}$ & $1.8\times10^{-5}$ \\
        $^{48}$Cr & $5.5\times10^{-4}$ & $2.1\times10^{-3}$ & $1.7\times10^{-3}$ & $2.3\times10^{-4}$ & $3.8\times10^{-4}$ & $3.6\times10^{-4}$ & $3.7\times10^{-4}$ & $3.9\times10^{-4}$ \\
        $^{52}$Fe & $2.0\times10^{-3}$ & $4.1\times10^{-3}$ & $6.5\times10^{-4}$ & $2.5\times10^{-5}$ & $8.7\times10^{-3}$ & $7.8\times10^{-3}$ & $8.1\times10^{-3}$ & $8.8\times10^{-3}$ \\
        $^{55}$Mn & $6.1\times10^{-8}$ & $5.9\times10^{-8}$ & $7.3\times10^{-8}$ & $9.9\times10^{-8}$ & $9.1\times10^{-8}$ & $4.4\times10^{-8}$ & $3.8\times10^{-8}$ & $4.4\times10^{-8}$ \\
        $^{55}$Co & $2.7\times10^{-4}$ & $4.8\times10^{-4}$ & $1.7\times10^{-5}$ & $1.5\times10^{-6}$ & $4.4\times10^{-3}$ & $4.0\times10^{-3}$ & $4.2\times10^{-3}$ & $4.5\times10^{-3}$ \\
        $^{56}$Ni & $3.9\times10^{-2}$ & $8.2\times10^{-3}$ & $6.0\times10^{-5}$ & $1.9\times10^{-6}$ & $7.2\times10^{-1}$ & $5.4\times10^{-1}$ & $5.9\times10^{-1}$ & $5.4\times10^{-1}$ \\ \hline
    \end{tabular}
\end{table*}

\begin{table}
    \caption{Final abundances for Model M11\_05.}
    \label{tab:abund11}
    \centering
    \begin{tabular}{@{}lrr}
        \hline
        & He detonation & core detonation \\
        & [$M_\odot$] & [$M_\odot$] \\ \hline
        $^4$He & $1.0\times10^{-2}$ & $8.4\times10^{-3}$ \\
        $^{12}$C & $5.7\times10^{-6}$ & $2.5\times10^{-6}$ \\
        $^{16}$O & $3.8\times10^{-3}$ & $7.5\times10^{-4}$ \\
        $^{28}$Si & $5.6\times10^{-2}$ & $4.6\times10^{-2}$ \\
        $^{32}$S & $2.4\times10^{-2}$ & $3.7\times10^{-2}$ \\
        $^{40}$Ca & $5.7\times10^{-3}$ & $1.0\times10^{-2}$ \\
        $^{44}$Ti & $1.6\times10^{-4}$ & $1.7\times10^{-5}$ \\
        $^{48}$Cr & $7.4\times10^{-4}$ & $3.2\times10^{-4}$ \\
        $^{52}$Fe & $2.1\times10^{-3}$ & $7.3\times10^{-3}$ \\
        $^{55}$Mn & $6.4\times10^{-8}$ & $5.7\times10^{-8}$ \\
        $^{55}$Co & $2.2\times10^{-4}$ & $3.7\times10^{-3}$ \\
        $^{56}$Ni & $1.2\times10^{-2}$ & $8.3\times10^{-1}$ \\ \hline
    \end{tabular}
\end{table}
The nucleosynthetic yields of the explosion models as determined in the
postprocessing step sensitively depend on the shell and core masses. The
abundances of $^4$He, $^{12}$C, $^{16}$O, $^{28}$Si, $^{32}$S, $^{40}$Ca,
$^{44}$Ti, $^{48}$Cr, $^{52}$Fe, $^{55}$Mn, $^{55}$Co and $^{56}$Ni at
$100\,\mathrm{s}$ after He ignition are given for all models in
Tables~\ref{tab:abund08}, \ref{tab:abund09}, \ref{tab:abund10}, and
\ref{tab:abund11}.  A detailed list of the nucleosynthesis yields can be found
in Appendix \ref{sec:aa}. In Tables~\ref{app:stab1_1} and \ref{app:stab2_1} we
list the final abundances of stable nuclides and radioactive nuclides with a
lifetime less than 2\,Gyr decayed to stability. Radioactive nuclides with a
longer lifetime are given with the yields at $t=100\,\mathrm{s}$. We further
list the yields of selected radioactive nuclides at $t=100\,\mathrm{s}$ in
Tables~\ref{app:rad1_1} and \ref{app:rad2_1}. The tables are created in the
same way as in \citet{seitenzahl2013a}. All abundances are given in solar
masses.

The abundances of the He shell detonation are dominated by IMEs. This is the
case as they are produced in low density regions. We note that the shell
detonation of models with the same shell mass, but different core mass, burn at
different densities. This then leads to a different abundance distribution
based on the individual density profile. In addition, the mixing of carbon into
the shell during the relaxation influences the yields. The additional carbon
isotopes stop burning following the $\alpha$-chain at an earlier point so that
less heavy elements are produced (see \citet{yoon2004b} and \citet{gronow2020a}
for details). This is the case because an enhanced carbon abundance leads to a
high $^{12}$C to $\alpha$ particle ratio. At the same time some carbon present
in the shell supports the production of heavy elements. Above a cross-over
temperature the $^{12}$C($\alpha$,~$\gamma$)$^{16}$O reaction is faster than
the triple-$\alpha$ reaction. Figure~5 in \citet{gronow2020a} shows the
cross-over temperature depending on the carbon enhancement.

Since observations disfavour a large amount of $^{56}$Ni and other heavy
elements such as titanium and chromium in the shell ejecta
\citep{hoeflich1996b, fink2010a, kromer2010a} only models with lower shell
masses seem to be suitable candidates for reproducing normal SNe~Ia. In our
study they are represented by M08\_03, M09\_03, M10\_03 and M10\_02. All four
models produce an amount between $9.9\times10^{-7}$\,$M_\odot$ (M08\_03) and
$6.0\times10^{-5}$\,$M_\odot$ (M10\_03) of $^{56}$Ni in the shell detonation.
The synthetic color light curves derived from the hydrodynamical simulation
models are usually too red. This is a feature in part caused by the amount of
$^{44}$\,Ti that is produced in the double detonation. In our models the total
amounts are between $2.3\times10^{-4}$\,$M_\odot$ (M08\_03) and
$2.7\times10^{-3}$\,$M_\odot$ (M08\_05). A comparison of all models shows that
more $^{56}$Ni is produced when the core mass and consequently the central
density increases, as is expected because then more material is available at
sufficiently high densities. The total $^{56}$Ni abundances of all models is in
the expected range of a SNe~Ia \citep[e.g.][]{stritzinger2006b}.

As mentioned in Section \ref{sec:IgnMech}, the shell of M08\_10\_r has the
composition and mass of the initial setup before the relaxation. This change
has an influence on the carbon detonation ignition. However, the
nucleosynthetic yields of the core detonation are very similar to M08\_10,
though less $^{12}$C is burned in M08\_10\_r which can be explained by the
larger core mass of M08\_10\_r due to the decreased amount of mixing. The shell
detonation of M08\_10\_r results in a lower production of IMEs as expected.
Instead more $^{56}$Ni is produced as the $\alpha$-chain is not stopped early
on by the presence of $^{12}$C. The $^{44}$Ti production is about the same in
both models, the influence the admixture has on the observables is therefore
only small and mostly due to $^{48}$Cr which is produced less in M08\_10\_r.

The final abundances of Model M10\_05 can be compared to M2a of
\citet{gronow2020a} since these have the same setup. While both models exhibit
the same carbon detonation ignition mechanism, the slightly more massive shell
and the solar metallicity of M10\_05 only have a small effect on the
nucleosynthesis yields. In both models about $0.01$\,$M_\odot$ of carbon and
oxygen are mixed into the shell. However, in our model the shell also contains
$^{14}$N and $^{22}$Ne representing metallicity with $^{22}$Ne being present
due to the mixing between core and shell. The production of IMEs in the shell
detonation is higher in M10\_05 as more $^{4}$He and $^{12}$C are burned.  The
abundance of $^{44}$Ti is of the same order of magnitude in both models. We
note though that about $0.04\,M_\odot$ less $^{56}$Ni are produced in total in
M10\_05.  Instead more stable nickel isotopes are produced
\citep{timmes2003a,kasen2009a,shen2018b}. The changes of the total $^{56}$Ni
and $^{58}$Ni abundances is due to the solar metallicity of M10\_05 compared to
zero metallicity of M2a. \citet{lach2020a} present a detailed discussion of the
nucleosynthesis yields of M10\_05 \citep[M2a$_\odot$ in][]{lach2020a} and in
part of M2a.

\begin{figure*}[h!]
    \centering
    \begin{subfigure}[c]{0.31\textwidth}
        \includegraphics[width=0.99\textwidth]{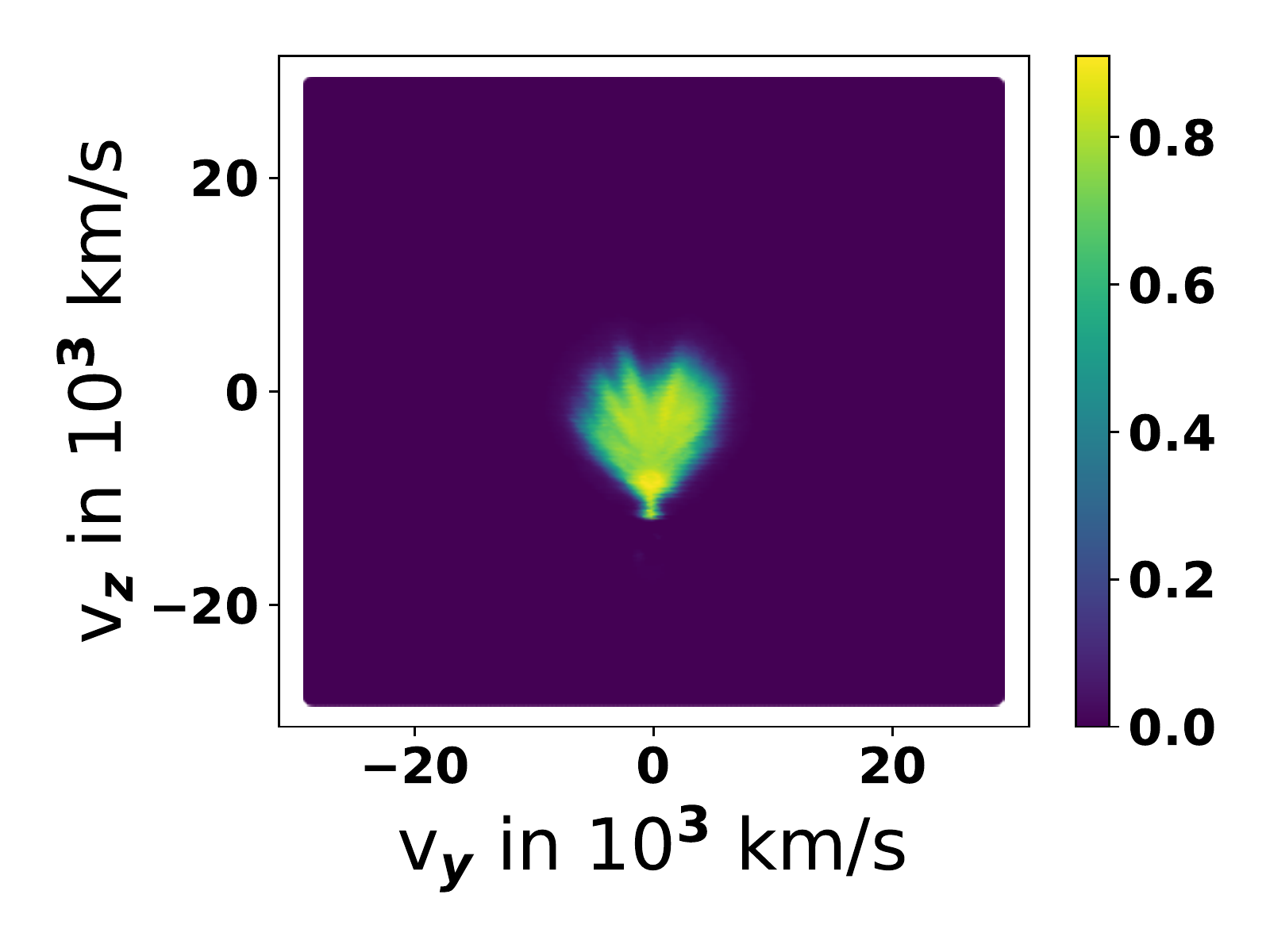}\subcaption{M08\_03}
    \end{subfigure}
    \begin{subfigure}[c]{0.31\textwidth}
        \includegraphics[width=0.99\textwidth]{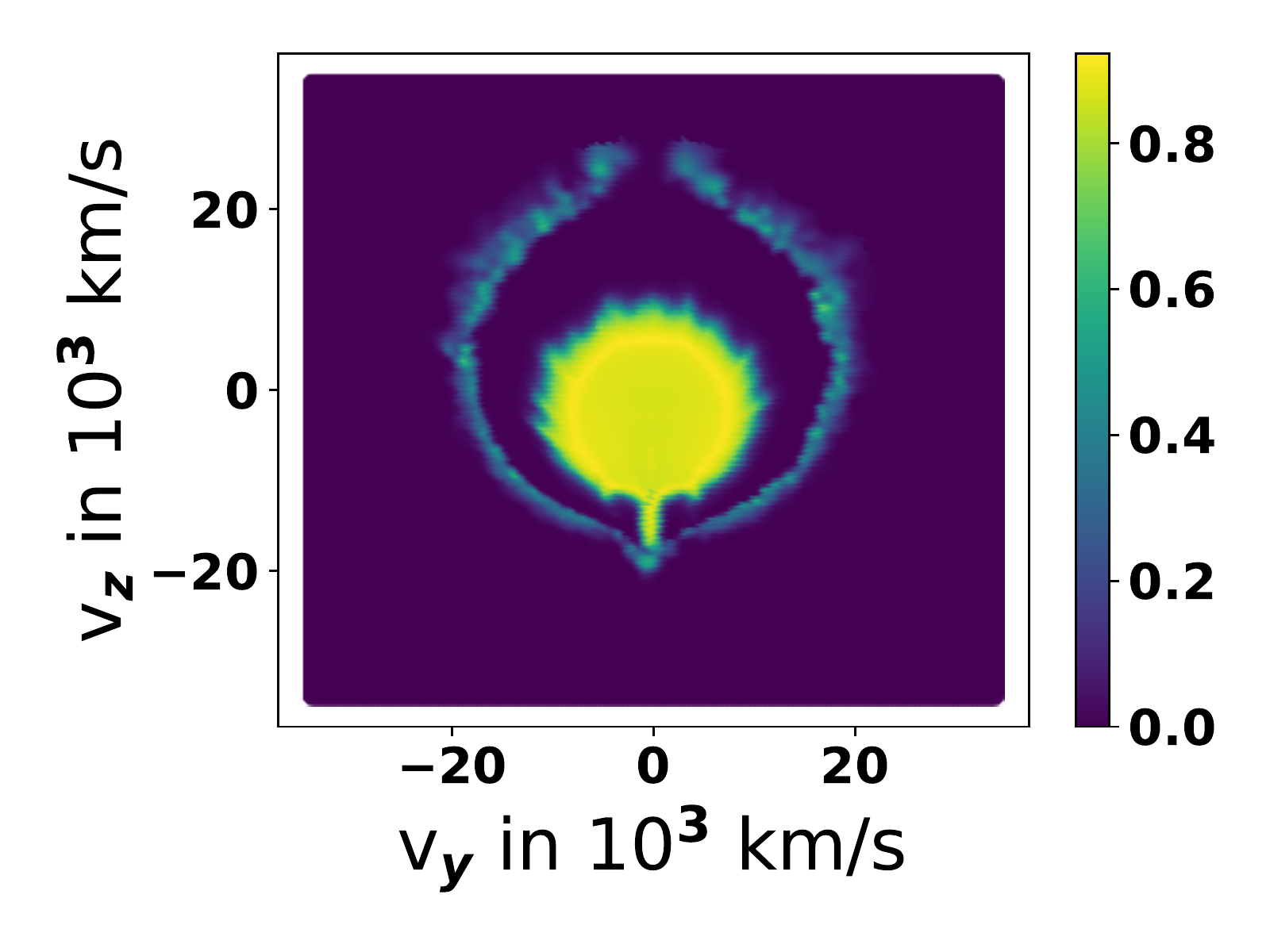}\subcaption{M10\_05}\label{m1005}
    \end{subfigure}
    \begin{subfigure}[c]{0.31\textwidth}
        \includegraphics[width=0.99\textwidth]{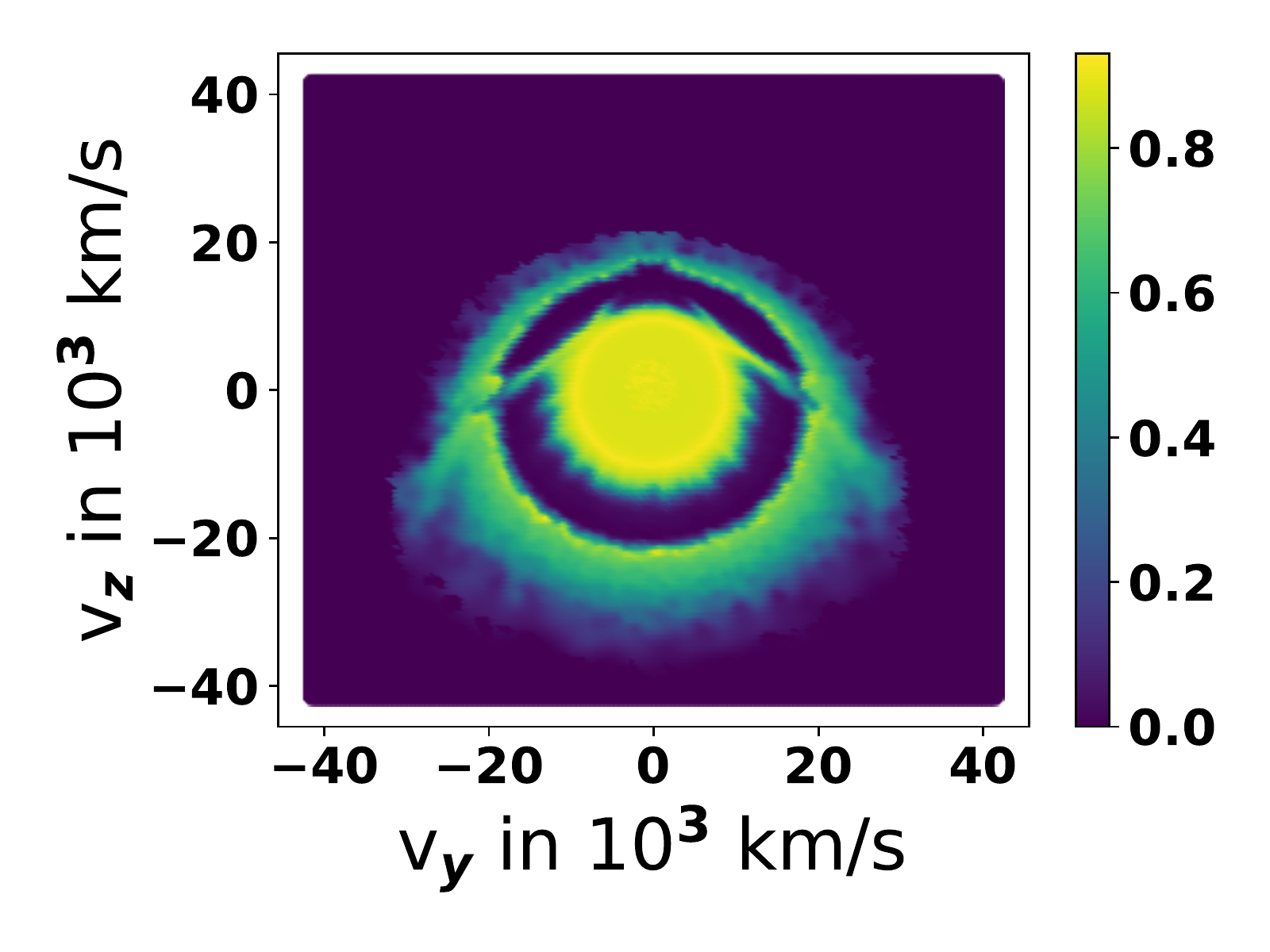}\subcaption{M10\_10}
    \end{subfigure}
    \caption{Slice along the x-axis of Models M08\_03 (left), M10\_05 (center)
    and M10\_10 (right) showing the $^{56}$Ni abundance in mass fraction for
the converging shock, scissors and edge-lit mechanism, respectively.}
    \label{fig:composition}
\end{figure*}

The spatial distribution of the nucleosynthesis yields in the ejecta
sensitively depends on the detonation mechanism of the double detonation
scenario. To illustrate this, Figure~\ref{fig:composition} shows the mass
fraction of $^{56}$Ni for Models M08\_03, M10\_05 and M10\_10, i.e.\ models
that exhibit the three different carbon ignition mechanisms. In M08\_03
$^{56}$Ni is only produced in the very center of the WD. The low core densities
only allow burning to $^{56}$Ni at shock convergence and at the center. The
temperature and density are too low for $^{56}$Ni production to take place in
other regions. Figure~\ref{m1005} shows a close resemblance to Figure~13 in
\citet{gronow2020a}. This is expected due to the similar setup of M2a and
M10\_05. In addition to the expected production of $^{56}$Ni in the core and to
some degree in the shell, M10\_05 also shows $^{56}$Ni at the convergence point
of the He detonation wave where carbon is ignited in the scissors mechanism. We
attribute this difference compared to M2a to the different density profiles.
The density at the convergence point of M10\_05 is higher than for M2a (maximum
values of $1.2\times10^7$\,$\text{g cm}^{-3}$ compared to
$8.3\times10^6$\,$\text{g cm}^{-3}$). The $^{56}$Ni production is more
symmetric for Model M10\_10. Here $^{56}$Ni is synthesized in the whole core as
well as in the shell. The larger $^{56}$Ni amount in the shell compared to
M10\_05 is due to the increased shell mass and thus higher density at the base
of the He shell. A small impact of the edge-lit detonation is visible as
$^{56}$Ni is located close to the base of the shell on the positive $z$-axis
and spread out more on the negative $z$-axis. The `wings' are a result of the
shallow transition region: He is burning in the whole transition region
propagating inwards. When it reaches the core, it causes a second detonation
wave to propagate through the core. However, as the core ignition is triggered
before by the edge-lit mechanism this does not have a big influence on the
abundances or evolution. A similar effect for M08\_10 and M09\_10 is discussed
in Section \ref{sec:IgnMech}. Its impact can potentially be decreased by
igniting a smaller region in the artificial He detonation or by a different
setup of the transition region. All models show a symmetry around the $z$-axis
because of the position of the He detonation spot we chose.

\section{Discussion in the context of other works}
\label{sec:discussion}
Simulations of exploding sub-M$_{\text{Ch}}$ WDs have previously been carried
out by other groups, such as \citet{livne1995a, garcia1999a, fink2010a,
moll2013a, blondin2017b, tanikawa2018a, polin2019a} and \citet{leung2020a}. A
comparison is difficult in some cases as most models do not account for
metallicity effects or the admixture of carbon into the He shell. Furthermore,
our simulations are among the first to be carried out in full 3D \citep[e.g.
see][]{moll2013a,tanikawa2018a}.

Since M10\_05 is set up in the same way as M2a in \citet{gronow2020a} a
comparison to Model 3 of \citealt{fink2010a} (FM3 hereafter) is also possible.
The model by \citet{fink2010a} assumes zero metallicity of the progenitor
material, and the shell in our model is more enriched in carbon. This carbon
enhancement has an influence on the abundances
\citep[see][]{kromer2010a,gronow2020a}. The differences in both the $^{44}$Ti
and the $^{48}$Cr yields from the shell detonation are, however, for the most
part caused by the different numerical treatments as can be seen from the
comparison of M1a and M2a with FM3 in \citet{gronow2020a}. The $^{44}$Ti and
$^{48}$Cr masses in our M10\_05 model are lower than in FM3 and match values of
M2a. The difference in the $^{44}$Ti mass is even one magnitude. Silicon and
sulfur are produced more in our model, which is due to the larger amount of
carbon in the helium shell, which causes the $\alpha$-chain to terminate at
IMEs. The differences in the yields from the shell detonation can in part be
attributed to the use of the level-set method for modeling the detonation
propagation assuming instantaneous burning of the fuel in \citet{fink2010a}. It
is not best suited to follow detonations at low densities, but is more accurate
at the high densities encountered in the core. In contrast to the
nucleosynthesis yields from the shell, the final yields from our core
detonation are more similar to that found by \cite{fink2010a}, with slight
differences resulting from the different setups and numerical treatments.

The model presented in \citet{tanikawa2018a} also has a similar setup with a
$0.95$\,$M_\odot$ core and a $0.05$\,$M_\odot$ shell. Their calculations are
carried out with an SPH code in 3D considering a small nuclear network of 13
species in the hydrodynamics. They acknowledge some mixing between core and
shell, but do not account for metallicity effects which is different to
M10\_05. We find a good agreement in the nucleosynthesis yields of the total
amount of produced $^{56}$Ni and lighter IMEs (Si and S). The amount of unburnt
oxygen matches as well. However, they find a slightly lower amount of $^{56}$Ni
produced in the shell detonation which is due to the different shell masses.

\citet{neunteufel2016a} argue that the binary systems considered in their study
cannot account for the majority of observed SNe~Ia. Their models accrete
$0.163$\,$M_\odot$ on average until detonation. The He shell should rather be
less massive as discussed by \citet{woosley2011b}. They find a set of models
that should result in normal SNe~Ia spectra \citep[see Table~1
in][]{neunteufel2016a}. \citet{woosley2011b} use the 1D Kepler code
\citep{weaver1978a,woosley2002a} to investigate different accretion rates and
luminosities in connection with varying WD core masses. They find that more
massive WDs have light curves like normal SNe~Ia when they have only a small He
shell. They further observe that less massive WDs accrete larger shells and
that hot WDs develop lower shell masses than cold WDs. The range of core masses
we consider in our study is also included in their parameter range. Out of the
models that show a double detonation (either edge-lit or converging shock),
seven models can be used for a comparison to our study. These are 10B,
10D, 8A, 10HB, 10HD, 9B and 8HBC \citep[for details see][]{woosley2011b}. A
detailed discussion of the nucleosynthesis yields of all models goes beyond
this work.  Instead we will focus on only some aspects in the following. It
needs to be considered that \citet{woosley2011b} do not account for metallicity
effects in their work. The total yields of $^{56}$Ni are about the same in
their models as in our models with a similar mass configuration (with two of
our models producing less). Among the selected models 10B and 10HD are both
similar to M10\_05. Both produce about the same amount of $^{28}$Si, $^{32}$S
and $^{40}$Ca. Our model produces significantly less $^{44}$Ti, $^{48}$Cr and
$^{52}$Fe than either of the two which indicates that the color may not be as
red as the ones of 10B and 10HD. However, the production of $^{55}$Co is higher
in our model having an influence on the final manganese yields after its decay
\citep[see][for a discussion on the effect manganese has on galactic chemical
evolution]{lach2020a}. Of the two models, 10B matches our model slightly better
as the discrepancy in the $^{56}$Ni production is not as big. Models 8A and
8HBC can be compared in the same way as they are similar to M08\_10\_r. Here 8A
and M08\_10\_r are a good match in the yields of $^{56}$Ni while we produce
more $^{44}$Ti, $^{48}$Cr and $^{55}$Co. On the other hand, 8HBC produces more
$^{44}$Ti and $^{56}$Ni. For a discussion of the difference between 8A and 8HBC
see \citet{woosley2011b}, we just mention that the initial luminosity is $100$
times higher in 8HBC representing a hot WD and that the IME production in both
models is almost the same. The differences between our models and the
comparison models of \citet{woosley2011b} are caused by the details of the
individual setups. Further, \citet{woosley2011b} only carry out 1D simulations.
Transferring their setup to 2D or 3D would result in a He detonation that
ignites in a shell and not just one point. The expansion of the material behind
the detonation front is therefore different from ours. However, we note that
their nuclear network includes up to about 500 species, with a base network of
19 isotopes, and adapts to the isotopes present in the time step. It should be
sufficient to gain reliable nucleosynthesis yields.

Another parameter study was carried out by \citet{polin2019a}. They consider a
set of core and shell masses in 1D simulations. Their models account for some
mixing taking place in a transition region between core and shell. Only the
radial span of the transition and not its composition are listed. Therefore a
detailed comparison with our models is difficult as we consider all cells with
an initial He mass fraction of at least $0.01$ as part of the shell. Their
models further do not include isotopes to represent metallicity. While
\citet{polin2019a} carry out a postprocessing step including a larger network,
their hydrodynamics simulations are considering only 21 isotopes. This low
number is not best suited for low shell masses as pointed out by
\citet{shen2014b} and \citet{townsley2019a}. Several of their models have a
good agreement with our models with respect to the core and shell masses. These
include models with a thin shell (a $1.0\,M_\odot$ with a $0.02\,M_\odot$
shell) as well as larger shells (such as a $0.8\,M_\odot$ or $0.9\,M_\odot$
core with a $0.08\,M_\odot$ shell, or a $1.0\,M_\odot$ with a $0.10\,M_\odot$
shell). In all models it is visible that the total IME yields are slightly
lower in our models, while in total more $^{56}$Ni is produced in some cases.
However, we observe that the $^{56}$Ni abundance coming from the shell
detonation is significantly lower in our models. Similar to
\citet{woosley2011b} this is most likely due to the different numerical
treatments as well as nuclear networks and the multi-dimensionality of our
study. \citet{polin2019a} state that their models with a thin He shell better
match some features of SNe Ia, such as characteristic spectral features. We
will discuss the spectral agreement of our models with observations in a
follow-up paper.

Model M10\_02 has a similar initial configuration as the model presented by
\citet{townsley2019a}. A detailed discussion is difficult as not all parameters
of their model are given. However, we note that the core is slightly less
massive in M10\_02 as material is mixed into the shell during the relaxation
step. The initial compositions of both models also show small differences.
However, both include $^{14}$N and $^{22}$Ne at solar metallicity. Our model
has a larger $^{56}$Ni production in the shell which is due to the higher
densities in the shell as this allows to burn to heavier elements. The
individual densities are in turn a result of the different initial temperature
profiles. At the same time, a larger amount of $^{44}$Ti from the shell
detonation in our model is caused by the stronger enrichment of carbon to the
shell. The nucleosynthesis yields following the core detonation show a very
good agreement. The discrepancy in the $^{56}$Ni yields can be explained by the
relatively high amount of elements with $Z\leq10$ in \citet{townsley2019a}.
More $^{4}$He is burned to higher-mass elements in M10\_02.

\section{Synthetic observables}
\label{sec:observables}

We have carried out three-dimensional radiative transfer calculations, using
the radiative transfer code \textsc{artis} \citep{kromer2009a, sim2007a} to
make predictions of the synthetic light curves for the explosion models. In
this paper we present the model bolometric light curves.  The model spectra at
a sequence of time steps are integrated over frequency to generate the
bolometric light curves. Each spectrum ranges from $600-30000 \ \AA$.  We will
discuss band-limited light curves, colours and spectra in a follow-up paper.

\subsection{Angle-averaged bolometric light curves.}

\begin{figure}[h!]
	\centering
	\includegraphics[width=0.45\textwidth]{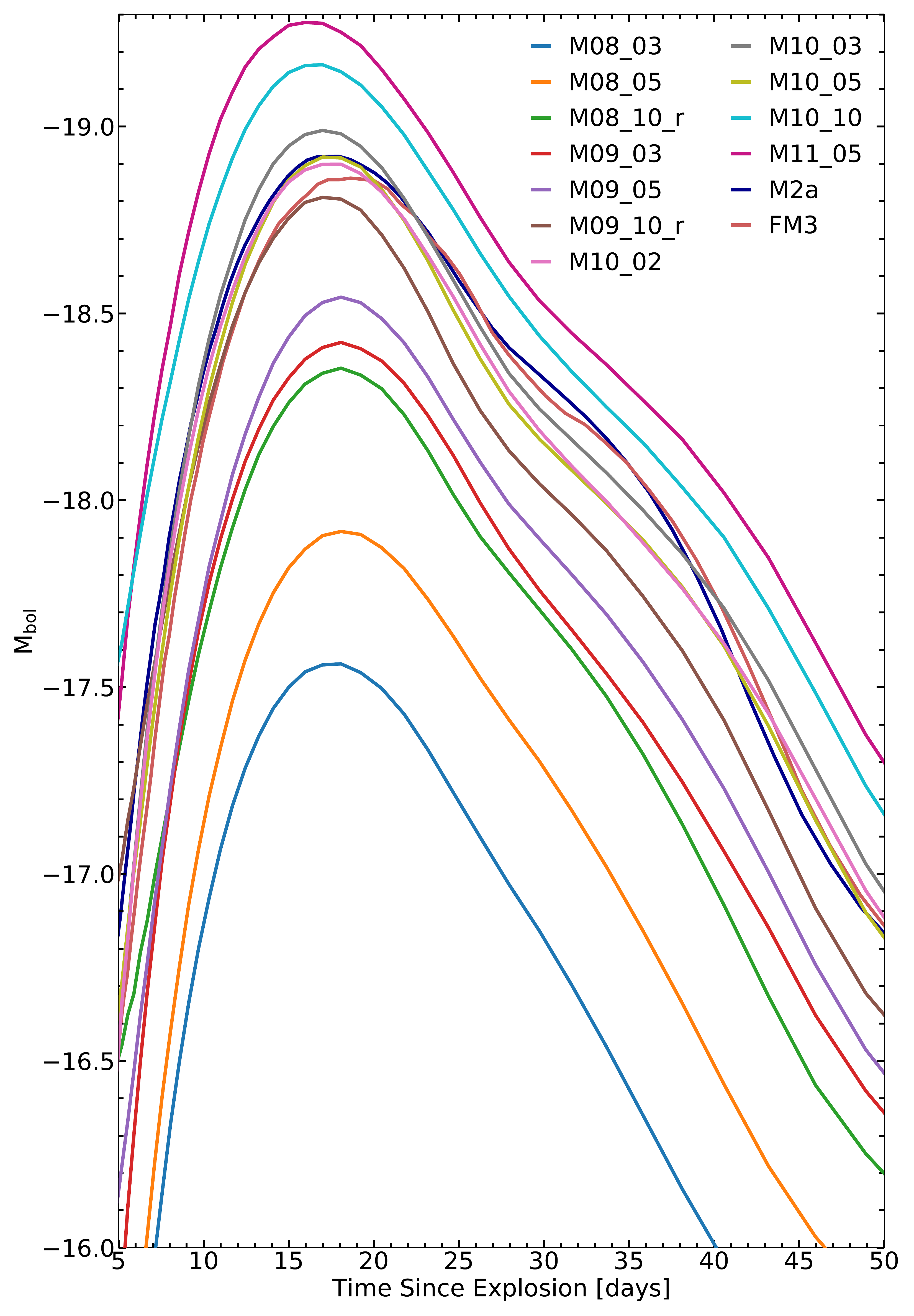}
   \caption{Angle-averaged bolometric light curves.
   For comparison we also show Model M2a from \cite{gronow2020a} and
   Model 3 from \citealt{fink2010a} (FM3).}
	\label{fig:bollightcurves}
\end{figure}

\begin{table*}[h]
    \tiny
    \caption{Parameters of the model angle-averaged bolometric light curves. The parameters for Model M2a \citep{gronow2020a} are also included for comparison.}
    \label{tab:paramstudy-observables}
    \centering
    \begin{tabular}{lllllllllllll}
        \hline
        & M08\_03 & M08\_05 & M08\_10\_r & M09\_03 & M09\_05 & M09\_10\_r & M10\_02 & M10\_03 & M10\_05 & M10\_10 & M11\_05 & M2a\\ \hline
        M$_{\mathrm{bol, max}}$         & -17.57   & -17.92   & -18.35   & -18.42   & -18.54   & -18.82   & -18.91   & -18.99   & -18.92   & -19.17 & -19.28 &  -18.93 \\
        t$_{\mathrm{bol, max}}$ [days] & 17.4    & 18.1    & 17.9    & 18.1    & 18.1    & 17.2    & 17.4    & 17.1    & 17.4    & 16.6  & 16.1 &  17.4 \\
        $\Delta \mathrm{m} _{15}$(bol)      & 0.93    & 0.85    & 0.83    & 0.86    & 0.81    & 0.89    & 0.86    & 0.86    & 0.88    & 0.82   & 0.81 & 0.71 \\
        mechanism                       & cs      & (s,) cs & s       & (s,) cs & (s,) cs & s       & art cs  & (s,) cs & s       & edge & edge & s \\ \hline
    \end{tabular}
\end{table*}

\begin{figure*}[h]
    \centering
    \includegraphics[width=0.99\textwidth]{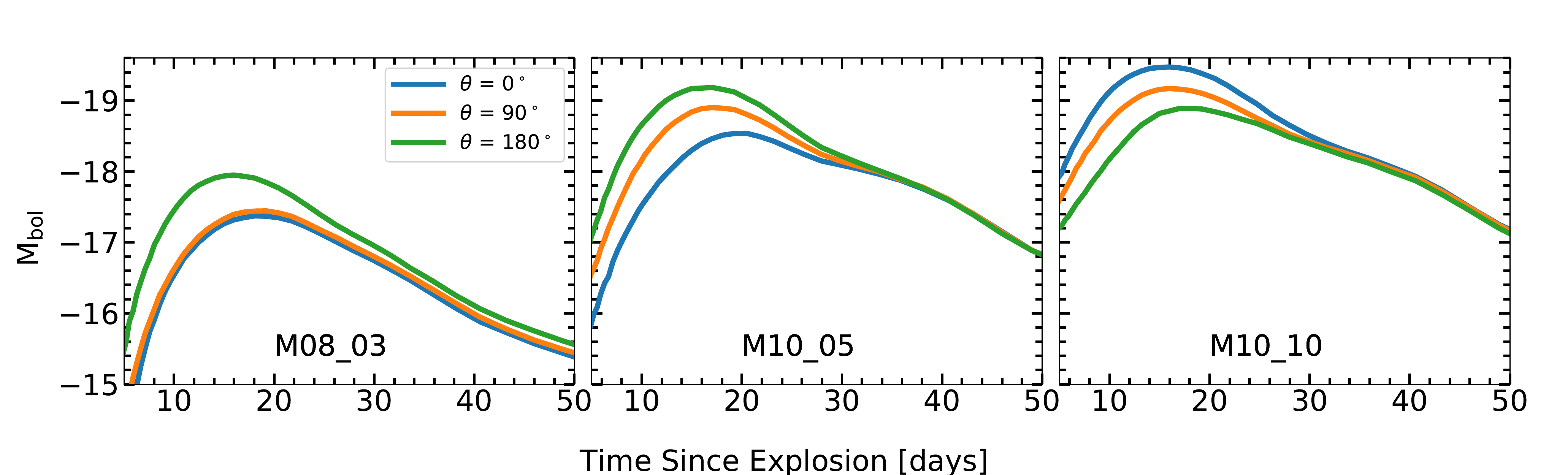}
    \caption{Bolometric line of sight dependent light curves for models M08\_03
    (converging shock), M10\_05 (scissors) and M10\_10 (edge-lit).}
    \label{fig:lightcurves-lineofsight}
\end{figure*}

In this section we present the angle-averaged light curves for each of the
models, and discuss the line of sight dependent light curves in
Sections~\ref{sec:lineofsight-bol-lc} and \ref{sec:width-luminosity}. Since the
models produce a range of masses of $^{56}$Ni, we expect the light curves to
show a range of brightnesses, increasing in brightness with the mass of
$^{56}$Ni. In Figure~\ref{fig:bollightcurves} we show the angle-averaged
bolometric light curves for each of the models. They demonstrate that the
models indeed produce light curves showing a range of brightnesses, accounting
for sub-luminous to normal brightness SNe Ia. The peak bolometric brightness
varies significantly between the models, and ranges from $-17.57\,\mathrm{mag}$
for our faintest model with the lowest mass of $^{56}$Ni, to
$-19.28\,\mathrm{mag}$ for the brightest model with the highest mass of
$^{56}$Ni (see Table~\ref{tab:paramstudy-observables}).  Apart from Model
M10\_05, the models show an increase in the mass of $^{56}$Ni with increasing
model mass. We therefore find that the peak brightness increases with model
mass, except for Model M10\_05 which is fainter than Model M10\_03. As the mass
of $^{56}$Ni produced in Model M10\_05 is less than Model M10\_03 this is
expected. However, this demonstrates that the light curve peak brightness is
not determined solely from the initial core and shell masses, and that the
detonation mechanism must be considered. We find that the peak brightness and
bolometric decline rate, $\Delta \mathrm{m} _{15}$(bol), shown by Model M10\_05
are similar to Model M10\_02.  These models produced the same mass of
$^{56}$Ni in the core.  The angle-averaged values of $\Delta \mathrm{m}
_{15}$(bol) are marked in Figure~\ref{fig:bolmax_deltam15-paramstudy}. We note
that Model M2a shows a very similar peak bolometric magnitude to Model M10\_05
($-18.93$ and $-18.92$, respectively). These models have the same initial core
and shell masses, and both show a secondary detonation by the scissors
mechanism, but Model M2a has zero metallicity (see Table~\ref{tab:model10} for
the metallicity of Model M10\_05). The decline rate $\Delta \mathrm{m}
_{15}$(bol) of Model M2a is, however, slower than Model M10\_05 ($0.71$
compared to $0.88$, see Table~\ref{tab:paramstudy-observables}). This indicates
that the model metallicity increases the rate of decline after maximum for this
progenitor configuration, but the peak bolometric brightness does not change.
As discussed in Section~\ref{sec:abundances}, the density at the base of the
shell is slightly lower in Model M10\_05 than in Model M2a, and this difference
in the density structure leads to a higher production of IMEs and lower
production of $^{56}$Ni in the shell detonation of Model M10\_05. It is likely
that the lower masses of heavy elements produced in the shell detonation when
metallicity is considered leads to lower opacities and hence a faster decline
rate. Model M2a shows a slower decline rate than all of the models in this
parameter study, which further supports that metallicity increases the
bolometric decline rate.

\subsection{Line of sight bolometric light curves}
\label{sec:lineofsight-bol-lc}
The double detonation scenario predicts an asymmetric distribution of $^{56}$Ni
in the explosion ejecta. This is shown in Figure~\ref{fig:composition} for
models exhibiting each of the three detonation scenarios found in this study
(converging shock, scissors, and edge-lit). For these three models (M08\_03,
M10\_05 and M10\_10) we show the light curves in the lines of sight viewing
towards the northern pole (i.e. in the direction of the initial He detonation,
$\theta = 0^\circ$), towards the equator ($\theta = 90^\circ$) and towards the
southern pole ($\theta = 180^\circ$) in
Figure~\ref{fig:lightcurves-lineofsight}. At maximum light, all of the models
show angle-dependent light curves, but over time the level of angle-dependence
decreases. As the ejecta become optically thinner with time, we find that the
dependence on the viewing angle decreases. For the M08\_03 and M10\_05 models
(showing secondary detonations by the converging shock, and scissors
mechanisms, respectively), the brightest line of sight is in the direction
towards the southern pole ($\theta = 180^\circ$). In the converging shock and
scissors scenarios, the line of sight at $\theta = 180^\circ$ has the highest
mass of $^{56}$Ni produced nearest to the surface of the ejecta, as can be seen
in Figure~\ref{fig:composition}. For the progenitor configuration considered
for Model M08\_03, the equatorial line of sight is similar to the line of sight
at $\theta = 0^\circ$ in bolometric light. This can also be explained by the
distribution of $^{56}$Ni, as similar amounts are produced in these lines of
sight. For a larger shell mass, however, the differences in these lines of
sight are greater (see Figure~\ref{fig:bolmax_deltam15-paramstudy}). In Model
M10\_05, the $^{56}$Ni synthesised in the core detonation is closer to the
surface of the ejecta in the equatorial line of sight ($\theta = 90^\circ$)
than in the line of sight at $\theta = 0^\circ$. Therefore the light curve at
$\theta = 90^\circ$ is brighter than at $\theta = 0^\circ$. In Model M10\_10,
however, the brightest line of sight is viewing towards the northern pole
($\theta = 0^\circ$). In the edge-lit scenario, the core detonation is ignited
on the same side of the core as the initial helium ignition. Therefore it is in
this line of sight that the $^{56}$Ni synthesised in the core detonation is
nearest to the ejecta surface (see Figure~\ref{fig:composition}, where the
outer ejecta are indicated by the outer ring of $^{56}$Ni produced in the shell
detonation).  The difference in brightness between the most extreme lines of
sight is similar to that found for Model M10\_05. Therefore we find that strong
asymmetries are expected in the light curves for the double detonation scenario
when considering the converging shock, the scissors and the edge-lit
mechanisms.

\subsection{Bolometric width-luminosity}
\label{sec:width-luminosity}
\begin{figure*}
    \centering
    \includegraphics[width=0.5\textwidth]{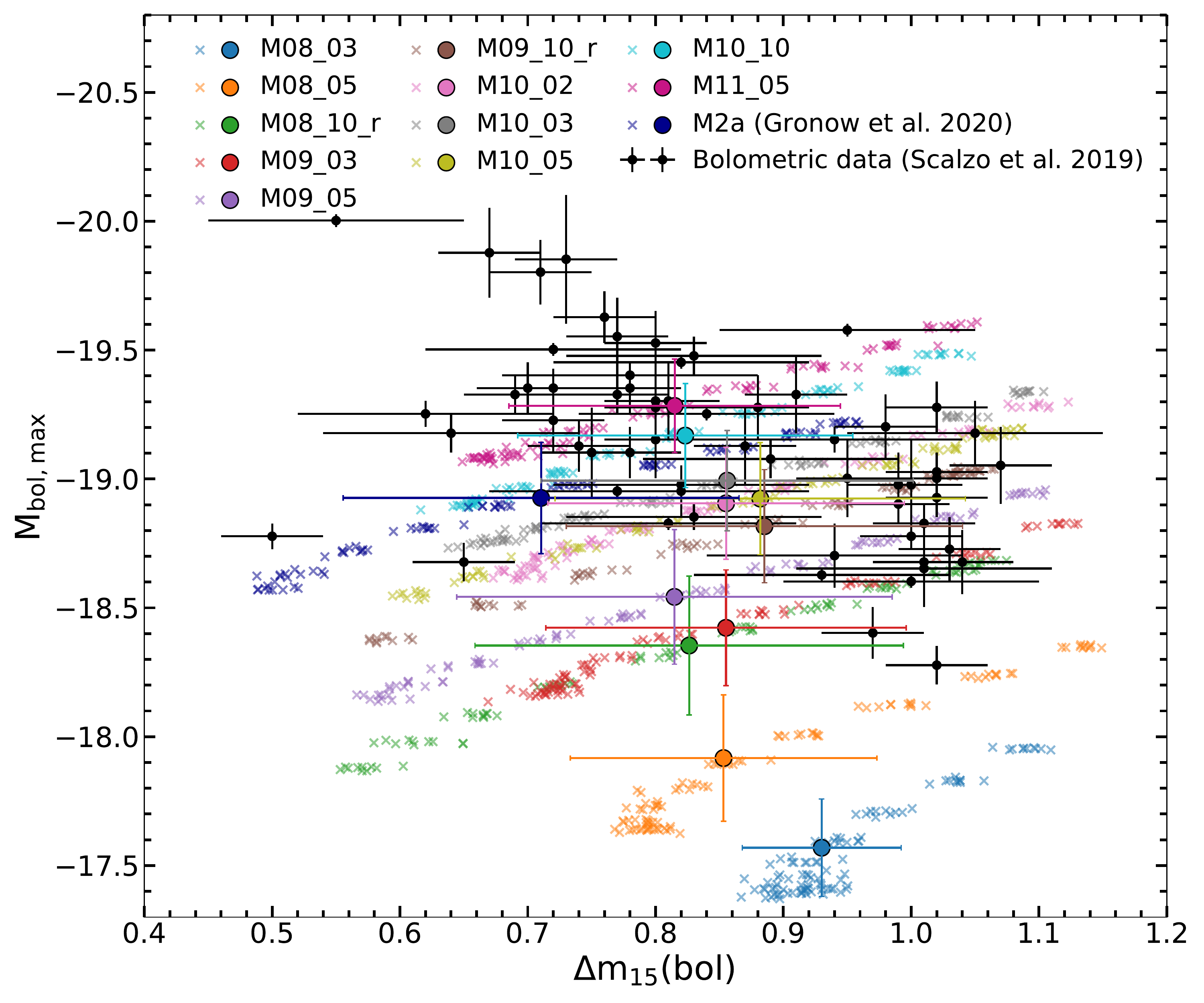}\includegraphics[width=0.5\textwidth]{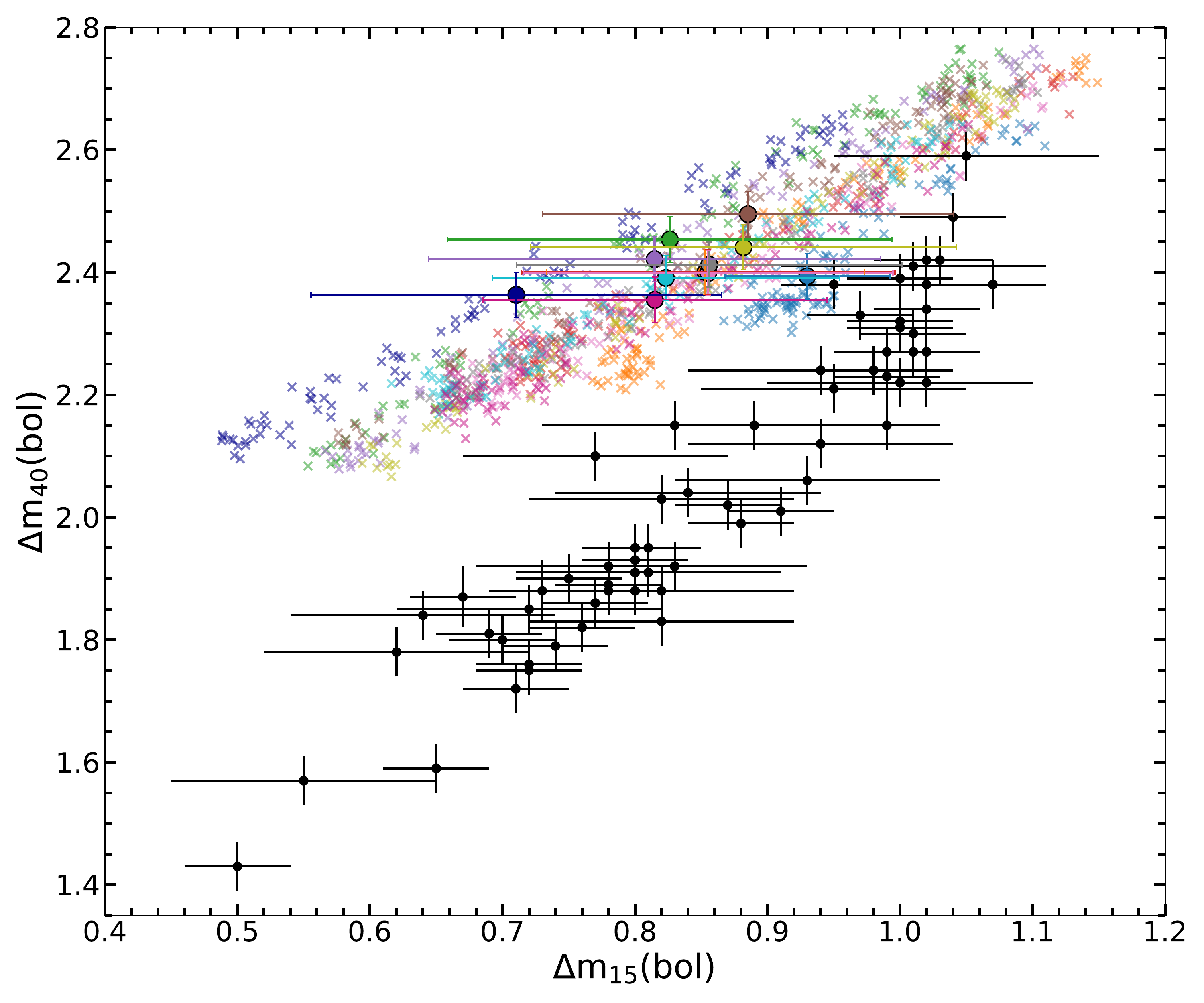}
    \caption{Left: peak bolometric magnitude plotted against
        $\Delta$m$_{15}$(bol).  Right: bolometric decline rate over 40 days,
        $\Delta\mathrm{m}_{40}$(bol), plotted against the decline rate over 15
        days, $\Delta\mathrm{m}_{15}$(bol).  We show the angle-averaged light
        curve values (circles) and 100 different viewing angles (x) for each
    model.  We include Model M2a \citep{gronow2020a} for comparison.  The error
bars show the standard deviation of the viewing angle distributions.  We also
plot the bolometric dataset of \cite{scalzo2019a}.}
    \label{fig:bolmax_deltam15-paramstudy}
\end{figure*}

To show the level of asymmetry in each of our models, we plot the maximum
bolometric magnitude against the bolometric decline rate over 15 days after
bolometric maximum, $\Delta\rm{m}_{15}$(bol), and show this for 100 different
viewing angles in each model (Figure~\ref{fig:bolmax_deltam15-paramstudy}). We
also show the angle-averaged values. It is clear that the angle-averaged values
do not well represent the true range of light curves produced by each of these
models. This highlights that the double detonation scenario is a fundamentally
multi-dimensional problem, and that to understand it multi-dimensional
simulations are required.

We find that within a model, the lines of sight tend to show an increasing
$\Delta \mathrm{m}_{15}$(bol) with increasing peak bolometric magnitude. In
brighter lines of sight the $^{56}$Ni is nearer the surface of the explosion
ejecta, leading to a brighter maximum, but then a faster decline from maximum.
In fainter lines of sight the $^{56}$Ni is deeper in the ejecta and the energy
takes longer to diffuse outwards, hence we find a fainter light curve that does
not fade as quickly.

The angle-averaged bolometric peak brightnesses increase with the total mass of
$^{56}$Ni produced by the model. However, the peak brightnesses in a given line
of sight do not necessarily increase with the total model mass of $^{56}$Ni, as
can be seen in Figure~\ref{fig:bolmax_deltam15-paramstudy}.  Model M2a shows
similar bolometric peak brightnesses to Model M10\_05, as expected given that
these models had the same initial masses, and both showed a secondary
detonation by the scissors mechanism. However, we find that for all lines of
sight, the bolometric decline rate for Model M2a is slower. Again, this
indicates that when the model metallicity is zero, we find a slower bolometric
decline from maximum. Due to the inclusion of $^{14}$N and $^{22}$Ne in Model
M10\_05, representing the model metallicity, the density at the base of the
shell is slightly lower in Model M10\_05 than in Model M2a.  The difference in
the density structure results in the higher production of IMEs and lower
production of heavy Fe-group elements in Model M10\_05. The higher abundance of
heavy Fe-group elements in Model M2a is likely responsible for the slower
decline rate found for Model M2a compared to Model M10\_05.

\subsection{Comparison to bolometric data}
In this section we make comparisons to the bolometric dataset constructed by
\cite{scalzo2019a} from well observed SNe~Ia. While band-limited light curves
provide much more accurate measurements of the data than constructing
bolometric light curves, band-limited light curves are challenging to simulate
since these require accurate calculations of the ejecta temperatures and
ionisation states, and are more sensitive to approximations made in the
radiative transfer calculations. Bolometric light curves show the total
radiant energy emitted as a function of time. This is dependent on the energy
deposition rate, primarily from the radioactive decays of $^{56}$Ni, and the
global opacity of the ejecta, and therefore is less sensitive to these
approximations. A comparison to bolometric data provides an excellent initial
test for how well our models agree with observations, while reducing the
uncertainties arising from approximations made in the radiative transfer
calculations. \cite{scalzo2019a} found a weak bolometric width-luminosity
relation. We plot the bolometric peak magnitudes and bolometric decline rates
$\Delta\rm{m}_{15}$(bol) from their sample in
Figure~\ref{fig:bolmax_deltam15-paramstudy}. Our brighter models lie in the
same parameter space as the bolometric data. However, the distribution shown by
the line of sight dependent values appears to be broader than the spread shown
by the bolometric data. The brighter models may loosely follow the observed
trend of the data, while our models extend to lower brightnesses and do not
account for the brighter SNe~Ia. The sample of bolometric data includes both
overluminous and subluminous SNe~Ia. The 1991bg-like SN~2006gt and SN~2007ba
are included in the sample, representing the faint end of observed SNe~Ia. Our
faintest model is $\sim 0.8$ mag fainter than the faintest SNe~Ia in that
sample, and all of the fainter models in our parameter study decline too slowly
compared to the data. Models M08\_03 and M08\_05 lie outside the parameter
space of the observed data, both in peak brightness and in decline rate. This
is similar to the findings of \cite{shen2018b}, who found their low mass 0.8
$\mathrm{M}_\odot$ pure detonation model (CO white dwarf without a He shell)
did not appear to match any observed SNe~Ia. \cite{shen2018b} suggested this
may be associated with a minimum white dwarf mass, following a discussion by
\cite{shen2014a} indicating that the central density may be too low to ignite a
secondary detonation by the converging shock mechanism. While we did find that
in Models M08\_03 and M08\_05 secondary detonations were ignited by the
converging shock mechanism, we also find that they do not resemble observations
of SNe~Ia in bolometric peak brightness and decline rate. This is also in
agreement with \cite{blondin2017a}, who found that their low mass (0.88
$\mathrm{M}_\odot$) pure detonation model showed an anti-width luminosity
relation.

\cite{scalzo2019a} found a strong correlation between $\Delta\rm{m}_{15}$(bol)
and $\Delta\rm{m}_{40}$(bol). In the right plot of
Figure~\ref{fig:bolmax_deltam15-paramstudy} we show this correlation for the
observed SNe~Ia, compared to our models. The angle-averaged model parameters
show a weak trend. However, the line of sight parameters show a clear
correlation between $\Delta\rm{m}_{15}$(bol) and $\Delta\rm{m}_{40}$(bol). The
models are, nevertheless, offset from the observed trend shown by the data.
The light curve behaviour at 40 days after maximum is driven by $\gamma$-ray
opacity, dependent on ejecta mass and velocity. Our models predict that the
bolometric light curves decline too quickly over 40 days compared to the data,
indicating that the optical depth is too low.  This implies that our model
masses may be too low, or our ejecta velocities or composition is not in
agreement with SNe~Ia. This is similar to the findings of \citet{kushnir2020b}
and \citet{sharon2020a} who showed that models fail to reproduce the observed
$\gamma$-ray escape time $t_0$ to $^{56}$Ni mass relation, and in particular,
luminous sub-M$_{\text{Ch}}$ models ($^{56}$Ni mass > 0.5 M$_\odot$) predict
values of $t_0 \approx 30$ days, which is lower than the observed SNe Ia values
of $t_0 = 35-45$ days. The models we consider in this work show similarly low
values of $t_0$. However, \citet{wygoda2019a} argue that sub-M$_{\text{Ch}}$
models are a better match to observations than M$_{\text{Ch}}$ models.

\section{Conclusions}
\label{sec:summary}
In this paper we presented full 3D simulations of double detonations in WDs
with varying core and He shell masses. The range of the study was chosen to
include low as well as high luminosity models. The core mass lies between
$0.8\,M_\odot$ and $1.1\,M_\odot$ while the shell has a mass between
$0.02\,M_\odot$ and $0.1\,M_\odot$. Looking into these different shell masses
allowed us to investigate their influence on the final yields coming from the
shell detonation. This is especially interesting as the He shell burning
products cause problems with the observables \citep{hoeflich1996b,kromer2010a}.
Not only the shell mass, but also the temperature profile and its effect on the
resulting density profile, the shell composition (especially C admixture which
suppresses the production of heavy elements and metallicity) and the
dimensionality of the model are important. A key parameter is also the method
used to follow the detonation, as well as the size of the network used in the
detonation simulations as pointed out by \citet{shen2018b} and
\cite{townsley2019a}. While the degree of mixing is an unknown
parameter and should be determined with better confidence, our models indicate
that the effect on the observables is low based on the small effect it has on
the nucleosynthetic abundances.

Most previous studies were not carried out in 3D, but 1D
\citep[e.g.][]{bildsten2007a, polin2019a} or 2D \citep[e.g.][]{fink2007a,
fink2010a, leung2020a}. Often the core and shell masses are chosen differently
from our models as well. In this context we note that the mass configuration of
core and shell in the detonation simulation is not set by us, but instead
partly a result of a numerical relaxation step. Our models match some of the
models of \cite{woosley2011b} and \citet{polin2019a} relatively well. This is
independent of whether the core detonation ignition is the same or not as in
\citet{woosley2011b}. Our models are potentially a better match to observations
than models by \citet{polin2019a} since less $^{56}$Ni is produced in the shell
detonation. A comparison to the models of \cite{woosley2011b} and
\citet{polin2019a} also shows the effect of multi-dimensional simulations and
metallicity.

The hydrodynamic treatment is different to previous work by
\citet{fink2007a,fink2010a} who use a level-set approach to follow the
detonation in a parameterized model, where the energy release in the burning
has to be calibrated. This introduces uncertainties. In our new models the
detonation is followed self-consistently by coupling a nuclear reaction network
to the hydrodynamics. Although the level-set approach has advantages when
modeling deflagration flames, our new approach seems more reliable when
following detonations, although they are not fully resolved and the detonation
structure is artificially broadened. Moreover, the ignition of the core
detonation arises spontaneously instead of being put in by hand. While
numerical effects may dominate the detonation ignition mechanism in our new
models, the thermodynamic conditions encountered in the ignition region justify
it in retrospect. Despite these differences and the superiority of the new
numerical treatment, the overall nucleosynthetic results of our new models
agree with the previous studies of \citet{fink2010a}. This shows that the
treatment of detonations with a calibrated level-set approach performs
surprisingly well. Our study finds varying carbon detonation ignition
mechanisms. They do not have a strong influence on the nucleosynthetic yields,
but they leave an imprint on the morphology of the ejecta.

The comparison of Model M10\_05 to M2a from \citet{gronow2020a} illustrates the
effect of metallicity on the final abundances in more detail. The
production of IMEs is higher in the shell detonation at solar metallicity as
the density profile is different. Furthermore, the core detonation produces more
stable isotopes of Ni. The $^{44}$Ti abundances do, however, not change
significantly.

A detailed discussion of the synthetic observables will be presented in a
follow-up paper. In this work we have discussed the bolometric properties of
our models. The explosion models produce a range of masses of $^{56}$Ni, and
therefore show a range of peak brightnesses, able to account for subluminous to
normal SNe~Ia brightnesses. The peak brightness of the models increases with
$^{56}$Ni mass, but it does not necessarily increase with the model mass.

Model M10\_05 and Model M2a have similar progenitor configurations. However,
Model M2a has zero metallicity. Comparing these models shows that the
consideration of the metallicity of the zero-age main sequence progenitor
impacts the rate of decline shown by the light curves. We found that Model M2a
has the slowest decline rate of all our models. This shows that metallicity
affects the predicted synthetic observables. Future work will investigate the
sensitivity to the chosen model metallicity.

The models again show that double detonations have a strong angle dependency.
This was previously discussed by \citet{gronow2020a}. We find that this is the
case for all of our models. All three core detonation mechanisms found in this
study (converging shock, scissors and edge-lit mechanism) produce highly
asymmetric ejecta profiles. Therefore when making predictions of the double
detonation scenario, multi-dimensional simulations are required. Angle-averaged
light curves are not a good representation of the true predicted observables in
a specific line of sight. The viewing angle effects on the bolometric light
curves are predominantly the result of the distribution of $^{56}$Ni in the
ejecta. Lines of sight where the $^{56}$Ni is closer to the ejecta surface show
brighter light curves.

We compared our parameter study models to the bolometric data set of
\cite{scalzo2019a}, and found that our brighter models lie in the same
parameter space as the fainter end of the observed bolometric width luminosity
relation. However, our faintest models are fainter than all SNe~Ia in the
sample. We also found that the distribution of the line of sight dependent
values shown by our models is broader than the sample of SNe~Ia. The bolometric
data set shows a strong correlation between $\Delta\mathrm{m}_{15}$(bol) and
$\Delta\mathrm{m}_{40}$(bol). Our models show a similarly strong correlation
when considering the line of sight specific values. However, we find that the
models are offset from the data, showing $\Delta\mathrm{m}_{40}$(bol) faster
than the SNe~Ia. Since the light curve behaviour at 40 days after maximum is
driven by $\gamma$-ray opacity, these results indicate that the model opacity
is lower than that of SNe~Ia, and suggest that the model masses may be too low,
or the ejecta velocities or composition is not in agreement with SNe~Ia. This
points to potentially generic problems of sub-Chandrasekhar mass WD explosions
as models for SNe~Ia, similar to those discussed by \citet{kushnir2020b} and
\citet{sharon2020a}. However, as illustrated by \citet{gronow2020a}, the models
remain promising for accounting for lower luminosity and/or spectroscopically
peculiar events. Further discussion of the comparison of our models to data,
including spectra and band light curves, will be presented in a companion paper
(Collins et al. in preparation).

\begin{acknowledgements}
This work was supported by the Deutsche Forschungsgemeinschaft (DFG, German
Research Foundation) -- Project-ID 138713538 -- SFB 881 (``The Milky Way
System'', subproject A10), by the ChETEC COST Action (CA16117), and by the
National Science Foundation under Grant No. OISE-1927130 (IReNA). SG and FKR
acknowledge support by the Klaus Tschira Foundation. SG thanks Florian Lach for
helpful discussions. NumPy and SciPy \citep{oliphant2007a}, IPython
\citep{perez2007a}, and Matplotlib \citep{hunter2007a} were used for data
processing and plotting. The authors gratefully acknowledge the Gauss Centre
for Supercomputing e.V. (www.gauss-centre.eu) for funding this project by
providing computing time on the GCS Supercomputer JUWELS \citep{juwels2019} at
J\"{u}lich Supercomputing Centre (JSC).
This work was performed using the Cambridge Service for Data Driven Discovery (CSD3),
part of which is operated by the University of Cambridge Research Computing on
behalf of the STFC DiRAC HPC Facility (www.dirac.ac.uk).
The DiRAC component of CSD3 was funded by BEIS capital funding via STFC
capital grants ST/P002307/1 and ST/R002452/1 and STFC operations
grant ST/R00689X/1. DiRAC is part of the National e-Infrastructure.
This research was undertaken with the assistance of resources from the
National Computational Infrastructure (NCI Australia), an NCRIS enabled capability
supported by the Australian Government.
\end{acknowledgements}

\bibliographystyle{aa} 

\begin{appendix}
\section{Abundances tables}
\label{sec:aa}

We list the nucleosynthesis yields of all our models in
Tables~\ref{app:stab1_1} and \ref{app:rad2_1}. Tables~\ref{app:stab1_1} to
\ref{app:stab2_1} list the stable nuclides, radioactive nuclides with lifetimes
lower than 2\,Gyr decayed to stability and radioactive nuclides with longer
lifetimes at time $t=100\,\mathrm{s}$. Nucleosynthesis yields of selected
radioactive nuclides at $t=100\,\mathrm{s}$ are listed in
Tables~\ref{app:rad1_1} and \ref{app:rad2_1}.
    \input{appendix_table_stable.tex}

\end{appendix}

\end{document}

%% file: appendix_table_stable.tex
\begin{landscape}
\small
\begin{table}
\centering
\caption{Asymptotic nucleosynthesis yields for Models M08\_10, M08\_05, M08\_03, M09\_10, M09\_05, and  M09\_03 (in solar masses).}
\label{app:stab1_1}
\begin{tabular}{l|cc|cc|cc|cc|cc|cc}
\hline
& \multicolumn{2}{c} {\text{M08\_10}} & \multicolumn{2}{c} {\text{M08\_05}} & \multicolumn{2}{c} {\text{M08\_03}} & \multicolumn{2}{c} {\text{M09\_10}} & \multicolumn{2}{c} {\text{M09\_05}} & \multicolumn{2}{c} {\text{M09\_03}} \\
 & He det & core det & He det & core det & He det & core det & He det & core det & He det & core det & He det & core det\\ 
 & [$M_\odot$] & [$M_\odot$] & [$M_\odot$] & [$M_\odot$] & [$M_\odot$] & [$M_\odot$] & [$M_\odot$] & [$M_\odot$] & [$M_\odot$] & [$M_\odot$] & [$M_\odot$] & [$M_\odot$]\\ \hline
$^{12}$C & 1.19e-04 & 1.05e-03 & 2.31e-03 & 7.46e-03 & 3.38e-03 & 1.25e-02 & 3.91e-05 & 1.32e-04 & 4.33e-04 & 2.63e-03 & 3.47e-03 & 4.88e-03\\
$^{13}$C & 5.81e-11 & 1.12e-10 & 4.08e-10 & 1.22e-07 & 1.41e-09 & 3.50e-07 & 1.69e-09 & 7.02e-12 & 1.98e-11 & 5.27e-09 & 8.25e-10 & 1.01e-07\\
$^{14}$N & 1.74e-05 & 1.73e-08 & 1.79e-05 & 7.46e-06 & 1.83e-05 & 1.02e-05 & 1.75e-05 & 2.17e-10 & 1.75e-05 & 6.02e-07 & 1.76e-05 & 3.94e-06\\
$^{15}$N & 1.86e-08 & 6.51e-10 & 6.27e-08 & 1.48e-08 & 4.42e-07 & 2.06e-08 & 2.59e-08 & 3.01e-10 & 1.83e-08 & 2.41e-09 & 1.04e-07 & 7.66e-09\\
$^{16}$O & 9.27e-03 & 8.08e-02 & 6.29e-03 & 1.16e-01 & 2.62e-03 & 1.45e-01 & 8.52e-03 & 5.50e-02 & 7.31e-03 & 7.78e-02 & 3.93e-03 & 9.22e-02\\
$^{17}$O & 1.10e-08 & 5.33e-09 & 3.02e-08 & 1.84e-06 & 6.01e-08 & 3.68e-06 & 1.10e-08 & 3.79e-11 & 1.14e-08 & 1.22e-07 & 3.99e-08 & 1.08e-06\\
$^{18}$O & 7.26e-08 & 2.52e-10 & 1.29e-07 & 3.27e-08 & 5.87e-07 & 4.98e-08 & 8.95e-08 & 2.01e-12 & 7.18e-08 & 3.73e-09 & 1.61e-07 & 1.62e-08\\
$^{19}$F & 9.88e-09 & 1.58e-11 & 4.05e-08 & 6.43e-10 & 5.06e-07 & 2.45e-09 & 1.13e-08 & 9.74e-13 & 1.18e-08 & 1.04e-10 & 8.06e-08 & 4.31e-10\\
$^{20}$Ne & 1.58e-04 & 2.86e-03 & 3.06e-03 & 5.23e-03 & 1.63e-03 & 6.24e-03 & 1.46e-05 & 6.24e-04 & 1.04e-03 & 3.56e-03 & 2.85e-03 & 3.32e-03\\
$^{21}$Ne & 4.19e-08 & 1.40e-07 & 2.64e-07 & 2.73e-06 & 1.21e-06 & 3.13e-06 & 5.23e-08 & 2.05e-08 & 5.49e-08 & 5.77e-07 & 4.12e-07 & 1.68e-06\\
$^{22}$Ne & 4.48e-07 & 3.45e-08 & 6.39e-07 & 7.21e-05 & 1.90e-06 & 2.67e-04 & 4.49e-07 & 5.40e-09 & 4.59e-07 & 2.69e-06 & 8.07e-07 & 5.20e-05\\
$^{23}$Na & 1.56e-06 & 1.62e-05 & 1.63e-05 & 7.22e-05 & 1.04e-05 & 9.07e-05 & 3.33e-07 & 3.77e-06 & 5.48e-06 & 2.76e-05 & 1.89e-05 & 4.41e-05\\
$^{24}$Mg & 3.16e-03 & 5.78e-03 & 4.21e-03 & 8.19e-03 & 2.09e-03 & 1.04e-02 & 2.48e-03 & 3.27e-03 & 3.25e-03 & 5.27e-03 & 3.17e-03 & 5.83e-03\\
$^{25}$Mg & 2.69e-06 & 3.44e-05 & 2.39e-05 & 1.43e-04 & 4.02e-05 & 1.71e-04 & 1.03e-06 & 9.04e-06 & 6.67e-06 & 5.75e-05 & 3.34e-05 & 8.72e-05\\
$^{26}$Mg & 2.60e-06 & 4.94e-05 & 3.72e-05 & 1.84e-04 & 3.55e-05 & 2.02e-04 & 9.01e-07 & 1.08e-05 & 1.02e-05 & 8.81e-05 & 4.53e-05 & 1.15e-04\\
$^{27}$Al & 7.90e-05 & 3.25e-04 & 1.49e-04 & 4.83e-04 & 8.26e-05 & 6.10e-04 & 3.79e-05 & 1.74e-04 & 1.10e-04 & 2.96e-04 & 1.13e-04 & 3.27e-04\\
$^{28}$Si & 1.28e-02 & 1.92e-01 & 9.25e-03 & 2.29e-01 & 4.21e-03 & 2.57e-01 & 1.26e-02 & 1.55e-01 & 9.97e-03 & 1.92e-01 & 5.77e-03 & 2.21e-01\\
$^{29}$Si & 9.09e-05 & 6.14e-04 & 1.11e-04 & 9.26e-04 & 5.34e-05 & 1.13e-03 & 6.87e-05 & 3.38e-04 & 9.01e-05 & 6.20e-04 & 8.14e-05 & 6.75e-04\\
$^{30}$Si & 1.07e-04 & 1.09e-03 & 1.07e-04 & 1.53e-03 & 6.99e-05 & 1.90e-03 & 9.22e-05 & 6.85e-04 & 9.71e-05 & 1.01e-03 & 8.29e-05 & 1.15e-03\\
$^{31}$P & 7.68e-05 & 4.53e-04 & 8.82e-05 & 6.18e-04 & 6.60e-05 & 7.58e-04 & 4.48e-05 & 3.07e-04 & 8.41e-05 & 4.17e-04 & 7.09e-05 & 4.79e-04\\
$^{32}$S & 5.51e-03 & 1.10e-01 & 4.69e-03 & 1.27e-01 & 2.40e-03 & 1.41e-01 & 4.28e-03 & 9.20e-02 & 4.42e-03 & 1.11e-01 & 2.76e-03 & 1.27e-01\\
$^{33}$S & 7.60e-05 & 3.14e-04 & 5.34e-05 & 4.17e-04 & 1.37e-05 & 5.07e-04 & 4.56e-05 & 2.24e-04 & 6.80e-05 & 2.86e-04 & 2.31e-05 & 3.36e-04\\
$^{34}$S & 1.31e-04 & 2.31e-03 & 2.57e-05 & 3.08e-03 & 1.14e-05 & 3.64e-03 & 2.21e-04 & 1.73e-03 & 5.56e-05 & 2.26e-03 & 1.22e-05 & 2.67e-03\\
$^{36}$S & 4.17e-09 & 1.49e-07 & 1.23e-08 & 3.52e-07 & 2.18e-08 & 4.03e-07 & 3.74e-09 & 7.02e-08 & 3.36e-09 & 1.90e-07 & 1.90e-08 & 2.44e-07\\
$^{35}$Cl & 3.31e-05 & 1.29e-04 & 4.85e-05 & 1.65e-04 & 4.91e-05 & 1.97e-04 & 1.95e-05 & 8.65e-05 & 3.92e-05 & 1.19e-04 & 4.14e-05 & 1.34e-04\\
$^{37}$Cl & 7.87e-06 & 2.35e-05 & 1.01e-05 & 3.00e-05 & 8.90e-07 & 3.48e-05 & 3.64e-06 & 1.81e-05 & 9.30e-06 & 2.30e-05 & 3.03e-06 & 2.71e-05\\
$^{36}$Ar & 1.81e-03 & 1.96e-02 & 2.11e-03 & 2.17e-02 & 1.11e-03 & 2.34e-02 & 1.24e-03 & 1.71e-02 & 1.53e-03 & 2.01e-02 & 1.13e-03 & 2.27e-02\\
$^{38}$Ar & 4.12e-05 & 1.00e-03 & 1.49e-05 & 1.31e-03 & 2.62e-06 & 1.52e-03 & 5.95e-05 & 7.74e-04 & 2.38e-05 & 1.02e-03 & 4.08e-06 & 1.20e-03\\
$^{40}$Ar & 6.85e-10 & 1.77e-08 & 9.59e-09 & 7.65e-08 & 2.04e-08 & 8.18e-08 & 2.76e-10 & 6.98e-09 & 1.32e-09 & 3.30e-08 & 1.65e-08 & 5.03e-08\\
$^{39}$K & 9.17e-05 & 6.52e-05 & 1.37e-04 & 8.22e-05 & 1.04e-04 & 9.39e-05 & 4.38e-05 & 5.11e-05 & 9.64e-05 & 6.53e-05 & 8.09e-05 & 7.61e-05\\
$^{41}$K & 3.57e-06 & 4.17e-06 & 9.55e-06 & 5.21e-06 & 5.35e-06 & 5.93e-06 & 1.40e-06 & 3.31e-06 & 6.57e-06 & 4.19e-06 & 1.04e-05 & 4.91e-06\\
$^{40}$Ca & 6.23e-03 & 1.74e-02 & 8.00e-03 & 1.85e-02 & 3.10e-03 & 1.94e-02 & 4.72e-03 & 1.58e-02 & 5.10e-03 & 1.83e-02 & 3.98e-03 & 2.04e-02\\
$^{42}$Ca & 7.70e-06 & 2.60e-05 & 6.18e-06 & 3.36e-05 & 7.83e-06 & 3.88e-05 & 4.08e-06 & 2.01e-05 & 8.51e-06 & 2.59e-05 & 5.50e-06 & 3.06e-05\\
$^{43}$Ca & 2.99e-05 & 1.77e-07 & 2.00e-05 & 1.58e-07 & 1.93e-05 & 1.83e-07 & 1.34e-05 & 2.89e-07 & 2.43e-05 & 1.83e-07 & 1.07e-05 & 1.28e-07\\
\hline
\end{tabular}
\end{table}
\end{landscape}

\begin{landscape}
\small
\begin{table}
\centering
\caption*{Table \ref{app:stab1_1} continued.}
\label{app:stab1_2}
\begin{tabular}{l|cc|cc|cc|cc|cc|cc}
\hline
& \multicolumn{2}{c} {\text{M08\_10}} & \multicolumn{2}{c} {\text{M08\_05}} & \multicolumn{2}{c} {\text{M08\_03}} & \multicolumn{2}{c} {\text{M09\_10}} & \multicolumn{2}{c} {\text{M09\_05}} & \multicolumn{2}{c} {\text{M09\_03}} \\
 & He det & core det & He det & core det & He det & core det & He det & core det & He det & core det & He det & core det\\ 
 & [$M_\odot$] & [$M_\odot$] & [$M_\odot$] & [$M_\odot$] & [$M_\odot$] & [$M_\odot$] & [$M_\odot$] & [$M_\odot$] & [$M_\odot$] & [$M_\odot$] & [$M_\odot$] & [$M_\odot$]\\ \hline
$^{44}$Ca & 1.79e-03 & 1.38e-05 & 2.68e-03 & 1.21e-05 & 2.16e-04 & 1.18e-05 & 8.85e-04 & 1.61e-05 & 2.03e-03 & 1.50e-05 & 7.20e-04 & 1.44e-05\\
$^{46}$Ca & 2.18e-11 & 5.18e-09 & 3.62e-09 & 3.14e-08 & 7.55e-09 & 3.27e-08 & 1.51e-11 & 4.97e-10 & 3.57e-10 & 1.54e-08 & 6.10e-09 & 2.10e-08\\
$^{48}$Ca & 6.30e-10 & 1.35e-10 & 9.97e-10 & 2.38e-09 & 1.84e-09 & 3.47e-09 & 6.28e-10 & 4.33e-12 & 6.53e-10 & 7.13e-10 & 1.29e-09 & 1.60e-09\\
$^{45}$Sc & 4.24e-06 & 2.49e-07 & 5.86e-06 & 3.29e-07 & 1.15e-06 & 3.71e-07 & 1.12e-06 & 1.93e-07 & 4.53e-06 & 2.52e-07 & 5.19e-06 & 2.88e-07\\
$^{46}$Ti & 7.74e-06 & 9.77e-06 & 3.51e-06 & 1.24e-05 & 1.14e-06 & 1.41e-05 & 3.61e-05 & 7.60e-06 & 4.43e-06 & 9.81e-06 & 2.79e-06 & 1.16e-05\\
$^{47}$Ti & 7.00e-05 & 6.12e-07 & 8.07e-05 & 5.99e-07 & 7.22e-06 & 6.66e-07 & 5.43e-05 & 7.72e-07 & 5.93e-05 & 6.37e-07 & 4.40e-05 & 5.64e-07\\
$^{48}$Ti & 3.82e-03 & 3.33e-04 & 2.59e-03 & 3.13e-04 & 7.37e-06 & 2.86e-04 & 1.92e-03 & 3.40e-04 & 4.60e-03 & 3.66e-04 & 1.06e-04 & 3.89e-04\\
$^{49}$Ti & 3.55e-05 & 2.41e-05 & 2.49e-05 & 2.35e-05 & 2.72e-07 & 2.20e-05 & 2.40e-05 & 2.36e-05 & 4.71e-05 & 2.60e-05 & 3.83e-06 & 2.80e-05\\
$^{50}$Ti & 1.02e-09 & 2.55e-08 & 8.32e-09 & 4.52e-08 & 1.89e-08 & 5.38e-08 & 9.27e-10 & 6.55e-09 & 1.91e-09 & 2.41e-08 & 1.62e-08 & 2.95e-08\\
$^{50}$V & 6.01e-10 & 2.20e-08 & 1.34e-09 & 3.15e-08 & 1.23e-09 & 3.90e-08 & 3.10e-10 & 9.97e-09 & 1.00e-09 & 2.03e-08 & 1.57e-09 & 2.16e-08\\
$^{51}$V & 2.96e-04 & 6.70e-05 & 1.19e-04 & 6.50e-05 & 2.64e-07 & 6.03e-05 & 2.57e-04 & 6.61e-05 & 3.91e-04 & 7.19e-05 & 5.09e-06 & 7.74e-05\\
$^{50}$Cr & 3.80e-05 & 2.27e-04 & 9.65e-06 & 2.50e-04 & 1.49e-07 & 2.62e-04 & 2.08e-04 & 2.00e-04 & 2.15e-05 & 2.33e-04 & 1.51e-06 & 2.64e-04\\
$^{52}$Cr & 7.48e-03 & 7.34e-03 & 8.05e-04 & 6.78e-03 & 1.14e-06 & 5.66e-03 & 3.99e-03 & 7.52e-03 & 5.15e-03 & 8.16e-03 & 4.41e-06 & 8.79e-03\\
$^{53}$Cr & 1.13e-04 & 7.01e-04 & 2.29e-05 & 6.62e-04 & 1.03e-07 & 5.67e-04 & 7.08e-05 & 7.06e-04 & 1.54e-04 & 7.69e-04 & 5.07e-07 & 8.30e-04\\
$^{54}$Cr & 5.56e-09 & 1.25e-07 & 4.06e-08 & 2.16e-07 & 8.08e-08 & 2.61e-07 & 4.30e-09 & 8.43e-08 & 1.04e-08 & 1.30e-07 & 8.43e-08 & 1.63e-07\\
$^{55}$Mn & 9.42e-04 & 3.86e-03 & 3.17e-05 & 3.58e-03 & 5.87e-07 & 2.91e-03 & 3.75e-04 & 3.96e-03 & 4.12e-04 & 4.24e-03 & 7.16e-07 & 4.59e-03\\
$^{54}$Fe & 7.33e-05 & 2.33e-02 & 1.08e-05 & 2.44e-02 & 1.27e-06 & 2.36e-02 & 5.02e-05 & 2.15e-02 & 9.12e-05 & 2.46e-02 & 1.01e-06 & 2.77e-02\\
$^{56}$Fe & 1.49e-02 & 3.12e-01 & 8.14e-05 & 2.01e-01 & 2.21e-05 & 1.32e-01 & 2.61e-02 & 4.77e-01 & 2.01e-03 & 3.84e-01 & 1.60e-05 & 3.30e-01\\
$^{57}$Fe & 1.34e-03 & 5.53e-03 & 1.24e-05 & 2.59e-03 & 3.50e-06 & 1.55e-03 & 2.47e-03 & 1.04e-02 & 1.44e-04 & 7.10e-03 & 3.93e-06 & 5.07e-03\\
$^{58}$Fe & 1.55e-07 & 9.49e-07 & 2.75e-06 & 4.03e-06 & 4.14e-06 & 5.21e-06 & 7.34e-08 & 4.70e-07 & 5.12e-07 & 1.38e-06 & 5.13e-06 & 2.41e-06\\
$^{59}$Co & 4.83e-05 & 1.33e-04 & 1.65e-05 & 1.90e-05 & 6.03e-06 & 1.06e-05 & 3.23e-04 & 3.68e-04 & 2.12e-05 & 1.92e-04 & 8.64e-06 & 7.76e-05\\
$^{58}$Ni & 2.12e-04 & 7.84e-03 & 1.89e-05 & 3.25e-03 & 5.52e-06 & 2.03e-03 & 6.59e-04 & 1.55e-02 & 1.00e-04 & 1.04e-02 & 7.40e-06 & 7.00e-03\\
$^{60}$Ni & 1.60e-03 & 1.98e-03 & 2.02e-05 & 2.17e-04 & 7.13e-06 & 1.22e-04 & 2.70e-03 & 5.73e-03 & 8.26e-05 & 2.89e-03 & 1.09e-05 & 9.51e-04\\
$^{61}$Ni & 3.05e-04 & 8.07e-05 & 7.62e-06 & 1.43e-05 & 2.48e-06 & 1.23e-05 & 3.97e-04 & 2.30e-04 & 1.36e-05 & 1.15e-04 & 2.91e-06 & 3.91e-05\\
$^{62}$Ni & 1.63e-04 & 6.73e-04 & 1.54e-05 & 1.07e-04 & 3.29e-06 & 8.82e-05 & 1.72e-04 & 1.92e-03 & 2.99e-05 & 9.71e-04 & 4.60e-06 & 3.35e-04\\
$^{64}$Ni & 6.62e-08 & 2.67e-06 & 2.46e-07 & 3.72e-06 & 2.25e-07 & 4.57e-06 & 9.61e-09 & 9.45e-07 & 1.83e-07 & 2.35e-06 & 2.98e-07 & 2.44e-06\\
$^{63}$Cu & 6.90e-06 & 4.82e-06 & 2.35e-06 & 7.33e-06 & 4.96e-07 & 8.82e-06 & 1.37e-05 & 2.65e-06 & 2.64e-06 & 5.16e-06 & 8.31e-07 & 5.02e-06\\
$^{64}$Zn & 1.35e-04 & 9.67e-06 & 2.11e-06 & 6.50e-06 & 1.28e-07 & 8.02e-06 & 2.70e-04 & 1.93e-05 & 4.54e-06 & 1.08e-05 & 2.48e-07 & 5.92e-06\\
$^{66}$Zn & 2.60e-05 & 3.18e-05 & 2.50e-06 & 2.92e-05 & 2.62e-07 & 3.58e-05 & 3.11e-05 & 4.90e-05 & 6.81e-06 & 3.23e-05 & 3.73e-07 & 2.41e-05\\
$^{67}$Zn & 4.92e-06 & 5.55e-07 & 9.79e-07 & 6.99e-07 & 3.56e-08 & 8.23e-07 & 2.45e-06 & 1.98e-07 & 2.63e-06 & 4.89e-07 & 6.03e-08 & 4.72e-07\\
$^{68}$Zn & 8.43e-06 & 2.24e-06 & 1.21e-06 & 2.99e-06 & 3.76e-08 & 3.70e-06 & 4.21e-06 & 1.41e-06 & 4.67e-06 & 1.64e-06 & 5.99e-08 & 2.01e-06\\
$^{70}$Zn & 2.65e-10 & 2.31e-08 & 2.93e-09 & 3.09e-08 & 4.30e-09 & 3.64e-08 & 8.05e-11 & 7.42e-09 & 1.15e-09 & 2.07e-08 & 5.23e-09 & 2.04e-08\\
$^{69}$Ga & 6.81e-07 & 9.18e-07 & 1.04e-07 & 1.17e-06 & 1.19e-08 & 1.40e-06 & 3.41e-07 & 6.68e-07 & 4.65e-07 & 6.68e-07 & 1.51e-08 & 8.02e-07\\
$^{71}$Ga & 5.87e-08 & 2.01e-07 & 2.90e-08 & 2.30e-07 & 8.17e-09 & 2.70e-07 & 2.83e-08 & 1.10e-07 & 5.84e-08 & 1.41e-07 & 1.01e-08 & 1.57e-07\\
\hline
\end{tabular}
\end{table}
\end{landscape}

\begin{landscape}
\small
\begin{table}
\centering
\caption{Asymptotic nucleosynthesis yields for Models M10\_10, M010\_05, M10\_03, M10\_02, and M11\_05 (in solar masses).}
\label{app:stab2_1}
\begin{tabular}{l|cc|cc|cc|cc|cc}
\hline
& \multicolumn{2}{c} {\text{M10\_10}} & \multicolumn{2}{c} {\text{M10\_05}} & \multicolumn{2}{c} {\text{M10\_03}} & \multicolumn{2}{c} {\text{M10\_02}} & \multicolumn{2}{c} {\text{M11\_05}} \\
 & He det & core det & He det & core det & He det & core det & He det & core det & He det & core det\\ 
 & [$M_\odot$] & [$M_\odot$] & [$M_\odot$] & [$M_\odot$] & [$M_\odot$] & [$M_\odot$] & [$M_\odot$] & [$M_\odot$] & [$M_\odot$] & [$M_\odot$]\\ \hline
$^{12}$C & 1.09e-05 & 1.65e-05 & 4.04e-05 & 4.36e-04 & 7.61e-04 & 1.23e-03 & 1.67e-03 & 1.95e-03 & 5.71e-06 & 2.48e-06\\
$^{13}$C & 3.87e-09 & 4.13e-12 & 1.35e-10 & 1.86e-10 & 4.22e-11 & 1.95e-09 & 2.79e-10 & 2.74e-08 & 8.96e-10 & 9.04e-12\\
$^{14}$N & 1.71e-05 & 2.86e-10 & 1.75e-05 & 3.63e-08 & 1.73e-05 & 1.78e-07 & 1.72e-05 & 1.00e-06 & 1.74e-05 & 9.72e-10\\
$^{15}$N & 5.58e-10 & 9.98e-10 & 8.73e-09 & 3.86e-10 & 2.32e-08 & 8.45e-10 & 4.86e-08 & 2.25e-09 & 4.52e-09 & 3.70e-09\\
$^{16}$O & 3.09e-03 & 2.73e-03 & 9.35e-03 & 6.08e-02 & 6.79e-03 & 4.88e-02 & 1.86e-03 & 5.70e-02 & 3.82e-03 & 7.53e-04\\
$^{17}$O & 1.08e-08 & 1.24e-12 & 1.10e-08 & 8.33e-09 & 1.21e-08 & 3.84e-08 & 1.84e-08 & 2.49e-07 & 1.10e-08 & 8.18e-12\\
$^{18}$O & 5.83e-08 & 5.00e-10 & 6.90e-08 & 2.98e-10 & 7.76e-08 & 1.10e-09 & 1.03e-07 & 4.29e-09 & 6.45e-08 & 2.14e-09\\
$^{19}$F & 2.88e-09 & 1.49e-11 & 5.44e-09 & 8.21e-12 & 1.50e-08 & 3.69e-11 & 3.85e-08 & 1.24e-10 & 3.99e-09 & 1.17e-12\\
$^{20}$Ne & 7.43e-06 & 1.46e-07 & 1.98e-05 & 1.50e-03 & 1.69e-03 & 1.78e-03 & 1.32e-03 & 1.90e-03 & 6.63e-06 & 4.34e-08\\
$^{21}$Ne & 2.37e-08 & 8.14e-11 & 2.73e-08 & 6.52e-08 & 7.84e-08 & 2.26e-07 & 1.57e-07 & 5.59e-07 & 2.37e-08 & 2.17e-13\\
$^{22}$Ne & 5.21e-07 & 6.42e-07 & 4.40e-07 & 2.37e-07 & 4.68e-07 & 9.12e-07 & 5.61e-07 & 1.15e-05 & 4.41e-07 & 1.77e-08\\
$^{23}$Na & 1.45e-07 & 2.27e-09 & 4.94e-07 & 8.54e-06 & 8.07e-06 & 1.29e-05 & 8.80e-06 & 1.84e-05 & 1.50e-07 & 5.29e-09\\
$^{24}$Mg & 2.51e-04 & 7.85e-05 & 2.94e-03 & 4.18e-03 & 3.49e-03 & 2.98e-03 & 1.52e-03 & 3.43e-03 & 3.28e-04 & 1.61e-05\\
$^{25}$Mg & 1.09e-06 & 1.55e-08 & 1.24e-06 & 1.92e-05 & 9.85e-06 & 2.65e-05 & 1.44e-05 & 3.74e-05 & 5.91e-07 & 3.11e-09\\
$^{26}$Mg & 1.02e-06 & 1.70e-08 & 1.05e-06 & 2.49e-05 & 1.65e-05 & 4.08e-05 & 2.01e-05 & 5.34e-05 & 6.69e-07 & 2.67e-09\\
$^{27}$Al & 2.57e-06 & 1.90e-06 & 5.18e-05 & 2.38e-04 & 1.27e-04 & 1.59e-04 & 5.57e-05 & 1.85e-04 & 3.03e-06 & 3.58e-07\\
$^{28}$Si & 3.70e-02 & 7.34e-02 & 1.31e-02 & 1.62e-01 & 8.87e-03 & 1.51e-01 & 2.93e-03 & 1.71e-01 & 5.58e-02 & 4.55e-02\\
$^{29}$Si & 3.52e-05 & 1.25e-05 & 8.07e-05 & 4.31e-04 & 9.28e-05 & 3.57e-04 & 4.17e-05 & 4.08e-04 & 4.88e-05 & 3.78e-06\\
$^{30}$Si & 5.51e-05 & 1.95e-05 & 1.01e-04 & 8.11e-04 & 9.96e-05 & 6.02e-04 & 4.34e-05 & 7.01e-04 & 6.69e-05 & 4.94e-06\\
$^{31}$P & 2.39e-05 & 1.30e-05 & 5.59e-05 & 3.57e-04 & 8.50e-05 & 2.60e-04 & 3.77e-05 & 3.00e-04 & 3.80e-05 & 4.40e-06\\
$^{32}$S & 1.59e-02 & 5.42e-02 & 4.89e-03 & 9.60e-02 & 3.68e-03 & 9.12e-02 & 1.60e-03 & 1.02e-01 & 2.44e-02 & 3.68e-02\\
$^{33}$S & 1.62e-05 & 1.32e-05 & 6.05e-05 & 2.46e-04 & 5.70e-05 & 1.85e-04 & 1.32e-05 & 2.12e-04 & 2.74e-05 & 5.22e-06\\
$^{34}$S & 1.51e-04 & 1.21e-04 & 1.69e-04 & 1.78e-03 & 2.91e-05 & 1.54e-03 & 9.85e-06 & 1.77e-03 & 1.89e-04 & 3.83e-05\\
$^{36}$S & 6.62e-10 & 7.62e-10 & 4.44e-09 & 9.76e-08 & 3.89e-09 & 9.88e-08 & 7.65e-09 & 1.29e-07 & 7.76e-10 & 1.59e-10\\
$^{35}$Cl & 8.81e-06 & 5.22e-06 & 1.67e-05 & 1.04e-04 & 3.07e-05 & 7.93e-05 & 2.57e-05 & 9.05e-05 & 1.07e-05 & 2.39e-06\\
$^{37}$Cl & 1.62e-06 & 2.18e-06 & 5.46e-06 & 1.91e-05 & 7.76e-06 & 1.70e-05 & 2.04e-06 & 1.89e-05 & 3.73e-06 & 1.15e-06\\
$^{36}$Ar & 2.78e-03 & 1.23e-02 & 1.38e-03 & 1.78e-02 & 1.21e-03 & 1.73e-02 & 7.88e-04 & 1.92e-02 & 4.31e-03 & 9.02e-03\\
$^{38}$Ar & 4.38e-05 & 6.85e-05 & 4.62e-05 & 8.01e-04 & 1.34e-05 & 7.42e-04 & 2.49e-06 & 8.32e-04 & 8.33e-05 & 2.78e-05\\
$^{40}$Ar & 1.31e-10 & 8.89e-11 & 3.50e-10 & 1.09e-08 & 2.05e-09 & 1.54e-08 & 6.89e-09 & 2.25e-08 & 1.78e-10 & 3.33e-11\\
$^{39}$K & 1.14e-05 & 6.02e-06 & 4.63e-05 & 5.60e-05 & 4.93e-05 & 5.00e-05 & 5.79e-05 & 5.51e-05 & 1.75e-05 & 3.69e-06\\
$^{41}$K & 3.87e-07 & 5.35e-07 & 2.08e-06 & 3.51e-06 & 6.35e-06 & 3.27e-06 & 4.73e-06 & 3.60e-06 & 1.13e-06 & 3.38e-07\\
$^{40}$Ca & 3.42e-03 & 1.34e-02 & 4.26e-03 & 1.65e-02 & 3.26e-03 & 1.63e-02 & 2.38e-03 & 1.81e-02 & 5.72e-03 & 1.04e-02\\
$^{42}$Ca & 1.03e-06 & 1.96e-06 & 5.25e-06 & 2.13e-05 & 3.38e-06 & 1.90e-05 & 2.45e-06 & 2.12e-05 & 2.52e-06 & 9.94e-07\\
$^{43}$Ca & 4.60e-06 & 3.14e-07 & 1.36e-05 & 7.10e-07 & 5.50e-06 & 3.15e-07 & 5.21e-06 & 2.70e-07 & 2.51e-06 & 4.11e-07\\
\hline
\end{tabular}
\end{table}
\end{landscape}

\begin{landscape}
\small
\begin{table}
\centering
\caption*{Table \ref{app:stab2_1} continued.}
\label{app:stab2_2}
\begin{tabular}{l|cc|cc|cc|cc|cc}
\hline
& \multicolumn{2}{c} {\text{M10\_10}} & \multicolumn{2}{c} {\text{M10\_05}} & \multicolumn{2}{c} {\text{M10\_03}} & \multicolumn{2}{c} {\text{M10\_02}} & \multicolumn{2}{c} {\text{M11\_05}} \\
 & He det & core det & He det & core det & He det & core det & He det & core det & He det & core det\\ 
 & [$M_\odot$] & [$M_\odot$] & [$M_\odot$] & [$M_\odot$] & [$M_\odot$] & [$M_\odot$] & [$M_\odot$] & [$M_\odot$] & [$M_\odot$] & [$M_\odot$]\\ \hline
$^{44}$Ca & 2.72e-04 & 1.79e-05 & 7.87e-04 & 2.11e-05 & 1.09e-03 & 1.78e-05 & 5.69e-04 & 1.77e-05 & 1.59e-04 & 1.74e-05\\
$^{46}$Ca & 1.28e-11 & 3.22e-13 & 1.48e-11 & 1.95e-09 & 8.30e-10 & 7.06e-09 & 2.85e-09 & 9.60e-09 & 1.33e-11 & 8.66e-14\\
$^{48}$Ca & 5.96e-10 & 3.47e-12 & 6.17e-10 & 3.12e-11 & 6.77e-10 & 3.00e-10 & 9.16e-10 & 5.81e-10 & 6.11e-10 & 7.17e-15\\
$^{45}$Sc & 3.09e-07 & 8.73e-08 & 2.47e-06 & 2.17e-07 & 3.31e-06 & 1.90e-07 & 2.59e-06 & 2.11e-07 & 5.62e-07 & 1.15e-07\\
$^{46}$Ti & 8.00e-06 & 1.11e-06 & 5.39e-06 & 7.98e-06 & 2.43e-06 & 7.43e-06 & 1.36e-06 & 8.21e-06 & 3.25e-06 & 6.85e-07\\
$^{47}$Ti & 1.68e-05 & 7.61e-07 & 3.20e-05 & 1.04e-06 & 4.01e-05 & 8.44e-07 & 3.06e-05 & 7.75e-07 & 5.99e-06 & 9.53e-07\\
$^{48}$Ti & 5.55e-04 & 3.81e-04 & 2.08e-03 & 3.59e-04 & 1.67e-03 & 3.67e-04 & 2.32e-04 & 3.94e-04 & 7.40e-04 & 3.24e-04\\
$^{49}$Ti & 9.97e-06 & 2.48e-05 & 2.33e-05 & 2.46e-05 & 1.31e-05 & 2.51e-05 & 4.26e-06 & 2.72e-05 & 1.53e-05 & 2.03e-05\\
$^{50}$Ti & 9.14e-10 & 2.13e-08 & 8.92e-10 & 9.15e-09 & 3.12e-09 & 1.21e-08 & 8.57e-09 & 1.52e-08 & 9.92e-10 & 1.31e-08\\
$^{50}$V & 2.24e-10 & 1.05e-10 & 3.60e-10 & 1.43e-08 & 1.19e-09 & 1.11e-08 & 7.80e-10 & 1.30e-08 & 4.05e-10 & 3.22e-11\\
$^{51}$V & 8.14e-05 & 6.74e-05 & 1.51e-04 & 6.85e-05 & 6.42e-05 & 6.99e-05 & 1.12e-05 & 7.54e-05 & 3.36e-05 & 5.54e-05\\
$^{50}$Cr & 9.37e-05 & 1.42e-04 & 2.34e-05 & 2.07e-04 & 4.66e-06 & 2.03e-04 & 1.61e-06 & 2.25e-04 & 4.47e-05 & 1.00e-04\\
$^{52}$Cr & 1.98e-03 & 8.66e-03 & 4.10e-03 & 7.82e-03 & 6.56e-04 & 8.13e-03 & 2.58e-05 & 8.82e-03 & 2.09e-03 & 7.31e-03\\
$^{53}$Cr & 6.03e-05 & 7.78e-04 & 6.72e-05 & 7.33e-04 & 1.36e-05 & 7.57e-04 & 1.48e-06 & 8.21e-04 & 7.76e-05 & 6.48e-04\\
$^{54}$Cr & 1.08e-08 & 2.34e-07 & 3.89e-09 & 9.16e-08 & 1.65e-08 & 8.53e-08 & 4.38e-08 & 9.85e-08 & 1.70e-08 & 1.45e-07\\
$^{55}$Mn & 2.69e-04 & 4.38e-03 & 4.85e-04 & 4.08e-03 & 1.74e-05 & 4.22e-03 & 1.67e-06 & 4.57e-03 & 2.28e-04 & 3.68e-03\\
$^{54}$Fe & 1.09e-03 & 1.90e-02 & 4.20e-05 & 2.23e-02 & 5.54e-06 & 2.23e-02 & 1.26e-06 & 2.48e-02 & 1.68e-03 & 1.46e-02\\
$^{56}$Fe & 3.93e-02 & 7.23e-01 & 8.25e-03 & 5.39e-01 & 6.99e-05 & 5.91e-01 & 1.36e-05 & 5.41e-01 & 1.20e-02 & 8.26e-01\\
$^{57}$Fe & 1.48e-03 & 1.70e-02 & 5.94e-04 & 1.21e-02 & 5.82e-06 & 1.33e-02 & 2.96e-06 & 1.13e-02 & 3.16e-04 & 2.10e-02\\
$^{58}$Fe & 2.52e-08 & 7.78e-07 & 7.79e-08 & 7.08e-07 & 1.03e-06 & 6.41e-07 & 2.57e-06 & 9.30e-07 & 2.92e-08 & 4.87e-07\\
$^{59}$Co & 3.69e-04 & 6.89e-04 & 2.91e-05 & 4.78e-04 & 9.46e-06 & 5.08e-04 & 5.40e-06 & 3.86e-04 & 4.13e-05 & 9.52e-04\\
$^{58}$Ni & 5.79e-04 & 2.55e-02 & 1.25e-04 & 1.83e-02 & 1.31e-05 & 2.05e-02 & 5.76e-06 & 1.70e-02 & 2.06e-04 & 3.04e-02\\
$^{60}$Ni & 2.06e-03 & 1.03e-02 & 7.57e-04 & 8.20e-03 & 1.64e-05 & 8.01e-03 & 5.65e-06 & 6.17e-03 & 2.67e-04 & 1.38e-02\\
$^{61}$Ni & 1.48e-04 & 4.03e-04 & 1.22e-04 & 3.19e-04 & 3.91e-06 & 3.19e-04 & 1.76e-06 & 2.43e-04 & 2.04e-05 & 5.12e-04\\
$^{62}$Ni & 9.92e-05 & 3.42e-03 & 8.10e-05 & 2.51e-03 & 7.01e-06 & 2.70e-03 & 2.87e-06 & 2.07e-03 & 2.41e-05 & 4.35e-03\\
$^{64}$Ni & 3.44e-09 & 2.18e-09 & 1.36e-08 & 1.54e-06 & 2.14e-07 & 1.21e-06 & 1.50e-07 & 1.43e-06 & 3.53e-09 & 9.96e-10\\
$^{63}$Cu & 1.50e-05 & 1.66e-06 & 5.38e-06 & 4.24e-06 & 1.76e-06 & 3.59e-06 & 5.14e-07 & 3.72e-06 & 8.06e-06 & 2.89e-06\\
$^{64}$Zn & 2.26e-04 & 2.97e-05 & 6.04e-05 & 2.69e-05 & 2.20e-06 & 2.39e-05 & 2.05e-07 & 1.93e-05 & 3.20e-05 & 4.46e-05\\
$^{66}$Zn & 1.64e-05 & 6.08e-05 & 1.31e-05 & 6.29e-05 & 1.05e-06 & 5.68e-05 & 2.46e-07 & 4.74e-05 & 3.19e-06 & 7.79e-05\\
$^{67}$Zn & 7.52e-07 & 4.15e-08 & 2.15e-06 & 3.30e-07 & 2.56e-07 & 2.79e-07 & 5.97e-08 & 3.15e-07 & 3.04e-07 & 5.47e-08\\
$^{68}$Zn & 1.03e-06 & 2.97e-08 & 3.55e-06 & 1.56e-06 & 3.16e-07 & 8.09e-07 & 5.84e-08 & 9.76e-07 & 3.43e-07 & 3.07e-08\\
$^{70}$Zn & 5.54e-11 & 5.12e-13 & 8.51e-11 & 1.21e-08 & 1.73e-09 & 1.02e-08 & 2.75e-09 & 1.17e-08 & 5.80e-11 & 3.38e-16\\
$^{69}$Ga & 6.82e-08 & 5.57e-10 & 3.00e-07 & 8.07e-07 & 5.82e-08 & 3.33e-07 & 9.76e-09 & 4.07e-07 & 2.51e-08 & 5.48e-11\\
$^{71}$Ga & 5.19e-09 & 2.61e-11 & 2.25e-08 & 1.49e-07 & 1.51e-08 & 6.96e-08 & 5.62e-09 & 8.58e-08 & 3.59e-09 & 2.19e-12\\
\hline
\end{tabular}
\end{table}
\end{landscape}

\begin{landscape}
\small
\begin{table}
\centering
\caption{Nucleosynthesis yields for select radioactive nuclides of Models M08\_10, M08\_05, M08\_03, M09\_10, M09\_05, and M09\_03 (in solar masses).}
\label{app:rad1_1}
\begin{tabular}{l|cc|cc|cc|cc|cc|cc}
\hline
& \multicolumn{2}{c} {\text{M08\_10}} & \multicolumn{2}{c} {\text{M08\_05}} & \multicolumn{2}{c} {\text{M08\_03}} & \multicolumn{2}{c} {\text{M09\_10}} & \multicolumn{2}{c} {\text{M09\_05}} & \multicolumn{2}{c} {\text{M09\_03}} \\
 & He det & core det & He det & core det & He det & core det & He det & core det & He det & core det & He det & core det\\ 
 & [$M_\odot$] & [$M_\odot$] & [$M_\odot$] & [$M_\odot$] & [$M_\odot$] & [$M_\odot$] & [$M_\odot$] & [$M_\odot$] & [$M_\odot$] & [$M_\odot$] & [$M_\odot$] & [$M_\odot$]\\ \hline
$^{14}$C & 6.14e-12 & 4.85e-09 & 1.94e-08 & 4.11e-06 & 7.71e-08 & 5.50e-06 & 1.09e-11 & 4.62e-13 & 5.06e-11 & 3.22e-07 & 3.63e-08 & 2.20e-06\\
$^{22}$Na & 1.33e-08 & 1.00e-08 & 8.39e-08 & 2.02e-08 & 9.95e-07 & 2.37e-08 & 1.31e-08 & 4.67e-09 & 2.20e-08 & 1.35e-08 & 1.63e-07 & 1.29e-08\\
$^{26}$Al & 6.81e-07 & 5.68e-06 & 1.00e-05 & 9.38e-06 & 1.33e-05 & 1.16e-05 & 1.18e-07 & 1.59e-06 & 3.04e-06 & 6.28e-06 & 1.37e-05 & 5.87e-06\\
$^{32}$Si & 2.45e-12 & 2.88e-10 & 3.20e-10 & 7.18e-09 & 2.33e-09 & 7.49e-09 & 2.26e-12 & 1.08e-10 & 4.36e-12 & 1.68e-09 & 9.71e-10 & 4.75e-09\\
$^{32}$P & 2.07e-08 & 2.91e-07 & 1.54e-08 & 4.73e-07 & 1.09e-08 & 5.68e-07 & 1.62e-08 & 1.71e-07 & 1.56e-08 & 2.95e-07 & 1.20e-08 & 3.42e-07\\
$^{33}$P & 1.65e-08 & 2.24e-07 & 6.81e-09 & 3.43e-07 & 6.61e-09 & 4.29e-07 & 1.30e-08 & 1.39e-07 & 1.05e-08 & 2.02e-07 & 4.83e-09 & 2.46e-07\\
$^{35}$S & 1.14e-08 & 3.48e-07 & 1.43e-08 & 5.19e-07 & 8.69e-09 & 6.27e-07 & 3.81e-09 & 1.67e-07 & 1.39e-08 & 3.33e-07 & 9.83e-09 & 3.64e-07\\
$^{36}$Cl & 7.68e-08 & 8.48e-07 & 2.88e-08 & 1.15e-06 & 6.12e-09 & 1.42e-06 & 6.33e-08 & 5.35e-07 & 5.00e-08 & 7.53e-07 & 1.19e-08 & 8.56e-07\\
$^{37}$Ar & 7.80e-06 & 2.28e-05 & 1.01e-05 & 2.89e-05 & 8.25e-07 & 3.34e-05 & 3.58e-06 & 1.76e-05 & 9.26e-06 & 2.24e-05 & 2.98e-06 & 2.62e-05\\
$^{39}$Ar & 5.98e-10 & 1.70e-08 & 1.17e-08 & 9.01e-08 & 2.74e-08 & 1.00e-07 & 1.67e-10 & 6.89e-09 & 1.21e-09 & 3.23e-08 & 2.15e-08 & 5.73e-08\\
$^{40}$K & 4.83e-09 & 8.20e-08 & 7.00e-09 & 1.13e-07 & 4.16e-09 & 1.36e-07 & 1.31e-09 & 4.36e-08 & 6.22e-09 & 7.47e-08 & 4.92e-09 & 8.02e-08\\
$^{41}$Ca & 3.57e-06 & 4.17e-06 & 9.55e-06 & 5.18e-06 & 5.34e-06 & 5.89e-06 & 1.40e-06 & 3.31e-06 & 6.57e-06 & 4.18e-06 & 1.04e-05 & 4.89e-06\\
$^{44}$Ti & 1.79e-03 & 1.37e-05 & 2.68e-03 & 1.19e-05 & 2.16e-04 & 1.16e-05 & 8.85e-04 & 1.60e-05 & 2.03e-03 & 1.48e-05 & 7.19e-04 & 1.43e-05\\
$^{48}$V & 1.21e-06 & 5.34e-08 & 1.21e-06 & 7.19e-08 & 1.10e-07 & 7.22e-08 & 5.36e-07 & 4.44e-08 & 2.12e-06 & 5.58e-08 & 5.76e-07 & 6.57e-08\\
$^{49}$V & 3.98e-07 & 2.96e-07 & 3.74e-07 & 3.69e-07 & 3.40e-08 & 4.27e-07 & 2.43e-07 & 2.24e-07 & 5.36e-07 & 2.83e-07 & 1.35e-07 & 3.25e-07\\
$^{48}$Cr & 3.82e-03 & 3.33e-04 & 2.59e-03 & 3.12e-04 & 7.23e-06 & 2.85e-04 & 1.92e-03 & 3.39e-04 & 4.60e-03 & 3.66e-04 & 1.05e-04 & 3.89e-04\\
$^{49}$Cr & 3.51e-05 & 2.38e-05 & 2.45e-05 & 2.31e-05 & 2.33e-07 & 2.15e-05 & 2.38e-05 & 2.34e-05 & 4.66e-05 & 2.57e-05 & 3.69e-06 & 2.76e-05\\
$^{51}$Cr & 2.61e-06 & 1.75e-06 & 1.35e-06 & 2.13e-06 & 1.04e-08 & 2.42e-06 & 2.03e-06 & 1.39e-06 & 3.69e-06 & 1.75e-06 & 9.57e-08 & 2.04e-06\\
$^{51}$Mn & 2.94e-04 & 6.52e-05 & 1.18e-04 & 6.28e-05 & 2.37e-07 & 5.78e-05 & 2.55e-04 & 6.47e-05 & 3.87e-04 & 7.01e-05 & 4.98e-06 & 7.53e-05\\
$^{52}$Mn & 9.81e-06 & 2.71e-06 & 2.06e-06 & 3.09e-06 & 9.67e-09 & 2.84e-06 & 4.12e-06 & 2.56e-06 & 1.12e-05 & 2.96e-06 & 1.17e-07 & 3.50e-06\\
$^{53}$Mn & 5.69e-06 & 2.75e-05 & 1.43e-06 & 3.17e-05 & 3.39e-08 & 3.27e-05 & 3.32e-06 & 2.45e-05 & 7.97e-06 & 2.90e-05 & 8.28e-08 & 3.42e-05\\
$^{54}$Mn & 2.66e-09 & 9.38e-08 & 9.85e-09 & 1.11e-07 & 6.18e-09 & 1.30e-07 & 1.84e-09 & 6.70e-08 & 3.96e-09 & 8.50e-08 & 1.71e-08 & 9.77e-08\\
$^{52}$Fe & 7.47e-03 & 7.31e-03 & 8.02e-04 & 6.75e-03 & 8.84e-07 & 5.63e-03 & 3.98e-03 & 7.50e-03 & 5.14e-03 & 8.13e-03 & 4.12e-06 & 8.76e-03\\
$^{53}$Fe & 1.07e-04 & 6.74e-04 & 2.14e-05 & 6.30e-04 & 3.33e-08 & 5.34e-04 & 6.74e-05 & 6.81e-04 & 1.46e-04 & 7.40e-04 & 3.99e-07 & 7.96e-04\\
$^{55}$Fe & 1.35e-06 & 6.10e-05 & 4.44e-07 & 7.52e-05 & 9.53e-08 & 8.62e-05 & 8.49e-07 & 4.69e-05 & 1.41e-06 & 6.13e-05 & 1.31e-07 & 7.20e-05\\
$^{59}$Fe & 4.29e-09 & 3.23e-07 & 9.78e-07 & 3.00e-06 & 2.50e-06 & 3.39e-06 & 9.31e-10 & 1.83e-08 & 1.10e-07 & 9.79e-07 & 1.75e-06 & 1.74e-06\\
$^{60}$Fe & 1.28e-08 & 2.37e-06 & 1.90e-06 & 1.06e-05 & 3.93e-06 & 1.14e-05 & 4.99e-10 & 2.90e-07 & 2.69e-07 & 5.75e-06 & 3.55e-06 & 6.96e-06\\
$^{55}$Co & 9.40e-04 & 3.80e-03 & 3.12e-05 & 3.50e-03 & 2.85e-07 & 2.82e-03 & 3.74e-04 & 3.91e-03 & 4.11e-04 & 4.18e-03 & 4.33e-07 & 4.52e-03\\
$^{56}$Co & 3.59e-06 & 1.31e-05 & 5.30e-07 & 1.33e-05 & 3.05e-08 & 1.09e-05 & 1.25e-06 & 1.41e-05 & 3.09e-06 & 1.47e-05 & 4.91e-08 & 1.56e-05\\
$^{57}$Co & 2.14e-06 & 8.08e-06 & 3.65e-06 & 9.72e-06 & 1.77e-06 & 1.09e-05 & 1.10e-06 & 6.69e-06 & 1.96e-06 & 8.28e-06 & 2.39e-06 & 9.52e-06\\
$^{58}$Co & 5.92e-08 & 6.11e-08 & 4.99e-07 & 7.83e-08 & 1.72e-07 & 9.95e-08 & 2.23e-08 & 4.02e-08 & 1.35e-07 & 5.35e-08 & 3.60e-07 & 5.90e-08\\
$^{60}$Co & 2.61e-08 & 1.36e-06 & 1.00e-06 & 2.68e-06 & 8.98e-07 & 3.28e-06 & 1.34e-09 & 3.07e-07 & 2.60e-07 & 1.90e-06 & 2.21e-06 & 1.72e-06\\
$^{56}$Ni & 1.49e-02 & 3.12e-01 & 6.67e-05 & 2.01e-01 & 9.94e-07 & 1.32e-01 & 2.61e-02 & 4.77e-01 & 1.99e-03 & 3.84e-01 & 1.01e-06 & 3.30e-01\\
$^{57}$Ni & 1.33e-03 & 5.52e-03 & 8.01e-06 & 2.58e-03 & 2.36e-07 & 1.54e-03 & 2.47e-03 & 1.04e-02 & 1.42e-04 & 7.09e-03 & 3.20e-07 & 5.06e-03\\
\hline
\end{tabular}
\end{table}
\end{landscape}

\begin{landscape}
\small
\begin{table}
\centering
\caption*{Table \ref{app:rad1_1} continued.}
\label{app:rad1_2}
\begin{tabular}{l|cc|cc|cc|cc|cc|cc}
\hline
& \multicolumn{2}{c} {\text{M08\_10}} & \multicolumn{2}{c} {\text{M08\_05}} & \multicolumn{2}{c} {\text{M08\_03}} & \multicolumn{2}{c} {\text{M09\_10}} & \multicolumn{2}{c} {\text{M09\_05}} & \multicolumn{2}{c} {\text{M09\_03}} \\
 & He det & core det & He det & core det & He det & core det & He det & core det & He det & core det & He det & core det\\ 
 & [$M_\odot$] & [$M_\odot$] & [$M_\odot$] & [$M_\odot$] & [$M_\odot$] & [$M_\odot$] & [$M_\odot$] & [$M_\odot$] & [$M_\odot$] & [$M_\odot$] & [$M_\odot$] & [$M_\odot$]\\ \hline
$^{59}$Ni & 1.34e-05 & 1.83e-05 & 8.62e-06 & 4.60e-06 & 9.89e-07 & 3.71e-06 & 6.92e-05 & 4.45e-05 & 9.19e-06 & 2.51e-05 & 2.31e-06 & 1.30e-05\\
$^{63}$Ni & 1.00e-08 & 1.22e-06 & 1.95e-07 & 2.48e-06 & 2.27e-07 & 2.75e-06 & 5.86e-10 & 2.06e-07 & 8.28e-08 & 1.80e-06 & 2.48e-07 & 1.66e-06\\
$^{62}$Zn & 1.59e-04 & 6.25e-04 & 1.32e-05 & 3.83e-05 & 2.39e-06 & 1.73e-06 & 1.68e-04 & 1.89e-03 & 2.69e-05 & 9.28e-04 & 3.39e-06 & 2.85e-04\\
$^{65}$Zn & 5.93e-07 & 7.73e-07 & 2.20e-07 & 9.29e-07 & 9.24e-09 & 1.17e-06 & 3.13e-07 & 5.07e-07 & 4.31e-07 & 5.58e-07 & 3.13e-08 & 6.35e-07\\
$^{65}$Ge & 3.28e-06 & 2.53e-07 & 1.21e-07 & 1.22e-08 & 1.06e-08 & 9.76e-10 & 4.25e-06 & 8.59e-07 & 4.18e-07 & 3.64e-07 & 9.86e-09 & 6.90e-08\\
\hline
\end{tabular}
\end{table}
\end{landscape}

\begin{landscape}
\small
\begin{table}
\centering
\caption{Nucleosynthesis yields for select radioactive nuclides for Models M10\_10, M10\_05, M10\_03, M10\_02, and M11\_05 (in solar masses).}
\label{app:rad2_1}
\begin{tabular}{l|cc|cc|cc|cc|cc}
\hline
& \multicolumn{2}{c} {\text{M10\_10}} & \multicolumn{2}{c} {\text{M10\_05}} & \multicolumn{2}{c} {\text{M10\_03}} & \multicolumn{2}{c} {\text{M10\_02}} & \multicolumn{2}{c} {\text{M11\_05}} \\
 & He det & core det & He det & core det & He det & core det & He det & core det & He det & core det\\ 
 & [$M_\odot$] & [$M_\odot$] & [$M_\odot$] & [$M_\odot$] & [$M_\odot$] & [$M_\odot$] & [$M_\odot$] & [$M_\odot$] & [$M_\odot$] & [$M_\odot$]\\ \hline
$^{14}$C & 3.28e-13 & 1.15e-15 & 3.94e-12 & 1.78e-08 & 7.86e-10 & 9.35e-08 & 9.14e-09 & 5.64e-07 & 2.64e-12 & 3.26e-16\\
$^{22}$Na & 8.03e-09 & 3.68e-09 & 5.29e-09 & 6.71e-09 & 3.05e-08 & 6.73e-09 & 8.07e-08 & 7.15e-09 & 6.13e-09 & 1.58e-08\\
$^{26}$Al & 6.03e-07 & 1.27e-10 & 1.13e-07 & 3.35e-06 & 5.02e-06 & 3.18e-06 & 6.95e-06 & 3.35e-06 & 2.23e-07 & 1.21e-10\\
$^{32}$Si & 5.03e-13 & 5.57e-13 & 2.80e-12 & 2.02e-10 & 1.10e-11 & 6.63e-10 & 2.77e-10 & 1.70e-09 & 6.21e-13 & 9.58e-14\\
$^{32}$P & 4.21e-09 & 3.04e-09 & 2.08e-08 & 2.15e-07 & 1.44e-08 & 1.66e-07 & 5.46e-09 & 1.99e-07 & 5.58e-09 & 6.89e-10\\
$^{33}$P & 1.75e-09 & 2.11e-09 & 1.68e-08 & 1.68e-07 & 8.37e-09 & 1.13e-07 & 1.87e-09 & 1.38e-07 & 2.11e-09 & 4.20e-10\\
$^{35}$S & 8.87e-10 & 1.56e-09 & 5.33e-09 & 2.49e-07 & 1.44e-08 & 1.85e-07 & 4.19e-09 & 2.24e-07 & 1.13e-09 & 3.73e-10\\
$^{36}$Cl & 8.43e-09 & 1.25e-08 & 7.85e-08 & 6.40e-07 & 3.97e-08 & 4.40e-07 & 6.06e-09 & 5.18e-07 & 1.30e-08 & 3.45e-09\\
$^{37}$Ar & 1.61e-06 & 2.17e-06 & 5.39e-06 & 1.86e-05 & 7.72e-06 & 1.66e-05 & 2.01e-06 & 1.85e-05 & 3.71e-06 & 1.14e-06\\
$^{39}$Ar & 4.68e-11 & 1.26e-10 & 2.32e-10 & 1.08e-08 & 1.90e-09 & 1.48e-08 & 8.16e-09 & 2.39e-08 & 7.60e-11 & 4.00e-11\\
$^{40}$K & 3.45e-10 & 7.03e-10 & 1.90e-09 & 6.21e-08 & 6.33e-09 & 4.22e-08 & 2.67e-09 & 5.00e-08 & 7.49e-10 & 2.79e-10\\
$^{41}$Ca & 3.86e-07 & 5.35e-07 & 2.08e-06 & 3.50e-06 & 6.35e-06 & 3.27e-06 & 4.72e-06 & 3.59e-06 & 1.13e-06 & 3.38e-07\\
$^{44}$Ti & 2.72e-04 & 1.79e-05 & 7.87e-04 & 2.10e-05 & 1.09e-03 & 1.77e-05 & 5.69e-04 & 1.76e-05 & 1.59e-04 & 1.74e-05\\
$^{48}$V & 1.65e-07 & 2.51e-08 & 5.88e-07 & 6.15e-08 & 5.47e-07 & 4.46e-08 & 3.38e-07 & 5.02e-08 & 1.28e-07 & 1.85e-08\\
$^{49}$V & 5.74e-08 & 1.04e-07 & 2.21e-07 & 2.51e-07 & 1.64e-07 & 2.11e-07 & 8.35e-08 & 2.37e-07 & 7.80e-08 & 7.53e-08\\
$^{48}$Cr & 5.54e-04 & 3.81e-04 & 2.08e-03 & 3.59e-04 & 1.66e-03 & 3.67e-04 & 2.32e-04 & 3.94e-04 & 7.39e-04 & 3.24e-04\\
$^{49}$Cr & 9.91e-06 & 2.46e-05 & 2.31e-05 & 2.43e-05 & 1.29e-05 & 2.49e-05 & 4.17e-06 & 2.69e-05 & 1.52e-05 & 2.03e-05\\
$^{51}$Cr & 4.42e-07 & 3.62e-07 & 1.13e-06 & 1.47e-06 & 5.51e-07 & 1.38e-06 & 1.25e-07 & 1.52e-06 & 3.60e-07 & 2.39e-07\\
$^{51}$Mn & 8.10e-05 & 6.70e-05 & 1.50e-04 & 6.70e-05 & 6.36e-05 & 6.85e-05 & 1.11e-05 & 7.39e-05 & 3.32e-05 & 5.52e-05\\
$^{52}$Mn & 1.25e-06 & 2.58e-06 & 4.71e-06 & 2.93e-06 & 9.90e-07 & 2.74e-06 & 1.88e-07 & 3.02e-06 & 1.01e-06 & 1.96e-06\\
$^{53}$Mn & 2.21e-06 & 1.71e-05 & 2.86e-06 & 2.58e-05 & 6.41e-07 & 2.57e-05 & 1.10e-07 & 2.80e-05 & 3.29e-06 & 1.26e-05\\
$^{54}$Mn & 8.54e-09 & 4.81e-09 & 1.61e-09 & 7.46e-08 & 5.06e-09 & 6.38e-08 & 8.39e-09 & 7.02e-08 & 1.46e-08 & 2.26e-09\\
$^{52}$Fe & 1.98e-03 & 8.66e-03 & 4.10e-03 & 7.80e-03 & 6.55e-04 & 8.11e-03 & 2.54e-05 & 8.80e-03 & 2.08e-03 & 7.30e-03\\
$^{53}$Fe & 5.80e-05 & 7.61e-04 & 6.44e-05 & 7.07e-04 & 1.30e-05 & 7.31e-04 & 1.35e-06 & 7.93e-04 & 7.43e-05 & 6.35e-04\\
$^{55}$Fe & 3.06e-06 & 8.04e-06 & 7.56e-07 & 5.00e-05 & 2.12e-07 & 4.81e-05 & 9.19e-08 & 5.27e-05 & 6.88e-06 & 4.48e-06\\
$^{59}$Fe & 3.98e-11 & 8.17e-11 & 4.98e-10 & 1.44e-07 & 2.63e-07 & 3.84e-07 & 8.76e-07 & 6.07e-07 & 1.82e-10 & 5.21e-11\\
$^{60}$Fe & 1.98e-12 & 3.90e-11 & 7.63e-10 & 1.10e-06 & 6.27e-07 & 2.72e-06 & 1.79e-06 & 3.30e-06 & 2.32e-11 & 2.51e-11\\
$^{55}$Co & 2.66e-04 & 4.38e-03 & 4.85e-04 & 4.03e-03 & 1.71e-05 & 4.17e-03 & 1.47e-06 & 4.52e-03 & 2.21e-04 & 3.68e-03\\
$^{56}$Co & 8.57e-07 & 1.60e-05 & 1.59e-06 & 1.93e-05 & 1.85e-07 & 1.61e-05 & 6.43e-08 & 1.67e-05 & 9.84e-07 & 1.46e-05\\
$^{57}$Co & 5.39e-07 & 1.92e-06 & 1.07e-06 & 7.34e-06 & 1.67e-06 & 7.10e-06 & 1.75e-06 & 7.51e-06 & 8.71e-07 & 1.62e-06\\
$^{58}$Co & 3.97e-09 & 1.67e-09 & 2.48e-08 & 5.22e-08 & 2.34e-07 & 3.46e-08 & 1.86e-07 & 3.86e-08 & 5.75e-09 & 7.42e-10\\
$^{60}$Co & 2.75e-11 & 8.99e-11 & 1.82e-09 & 7.75e-07 & 5.09e-07 & 9.98e-07 & 1.04e-06 & 1.03e-06 & 7.02e-11 & 5.83e-11\\
$^{56}$Ni & 3.93e-02 & 7.23e-01 & 8.23e-03 & 5.38e-01 & 5.97e-05 & 5.91e-01 & 1.87e-06 & 5.41e-01 & 1.20e-02 & 8.26e-01\\
$^{57}$Ni & 1.48e-03 & 1.70e-02 & 5.93e-04 & 1.21e-02 & 3.77e-06 & 1.33e-02 & 4.81e-07 & 1.13e-02 & 3.15e-04 & 2.10e-02\\
\hline
\end{tabular}
\end{table}
\end{landscape}

\begin{landscape}
\small
\begin{table}
\centering
\caption*{Table \ref{app:rad2_1} continued.}
\label{app:rad2_2}
\begin{tabular}{l|cc|cc|cc|cc|cc}
\hline
& \multicolumn{2}{c} {\text{M10\_10}} & \multicolumn{2}{c} {\text{M10\_05}} & \multicolumn{2}{c} {\text{M10\_03}} & \multicolumn{2}{c} {\text{M10\_02}} & \multicolumn{2}{c} {\text{M11\_05}} \\
 & He det & core det & He det & core det & He det & core det & He det & core det & He det & core det\\ 
 & [$M_\odot$] & [$M_\odot$] & [$M_\odot$] & [$M_\odot$] & [$M_\odot$] & [$M_\odot$] & [$M_\odot$] & [$M_\odot$] & [$M_\odot$] & [$M_\odot$]\\ \hline
$^{59}$Ni & 4.28e-05 & 8.10e-05 & 6.80e-06 & 5.79e-05 & 6.23e-06 & 6.11e-05 & 1.54e-06 & 4.68e-05 & 3.58e-06 & 1.04e-04\\
$^{63}$Ni & 1.70e-11 & 1.76e-10 & 9.19e-10 & 5.27e-07 & 1.29e-07 & 9.05e-07 & 1.26e-07 & 9.72e-07 & 3.80e-11 & 1.09e-10\\
$^{62}$Zn & 9.89e-05 & 3.42e-03 & 7.65e-05 & 2.47e-03 & 4.42e-06 & 2.68e-03 & 2.25e-06 & 2.04e-03 & 2.39e-05 & 4.35e-03\\
$^{65}$Zn & 2.97e-08 & 1.25e-08 & 1.81e-07 & 7.04e-07 & 3.26e-07 & 3.02e-07 & 1.86e-08 & 3.69e-07 & 6.13e-09 & 1.43e-08\\
$^{65}$Ge & 3.24e-06 & 1.47e-06 & 1.72e-06 & 1.33e-06 & 3.98e-08 & 1.16e-06 & 9.83e-09 & 8.86e-07 & 8.24e-07 & 2.03e-06\\
\hline
\end{tabular}
\end{table}
\end{landscape}